\documentclass{my_elsart} 
\usepackage{amsmath}
\usepackage{my_SIunits}  
\usepackage{psfrag}
\usepackage{natbib}
\usepackage[dvipdfm]{graphicx}
\usepackage[latin1]{inputenc} 
\usepackage[english]{babel} 
\usepackage{moreverb}       

\newcommand{\xh}[0]{\hat{X}}
\newcommand{\ph}[0]{\hat{P}}

\renewcommand{\S}[1]{\hat{S}_{#1}}
\renewcommand{\a}[1]{{\hat{a}}_{#1}}

\newcommand{\X}[1]{\hat{X}_{#1}}
\renewcommand{\P}[1]{\hat{P}_{#1}}

\newcommand{\adag}[1]{\hat{a}^{\dagger}_{#1}}

\newcommand{\J}[1]{{\hat{J}}_{#1}}
\renewcommand{\j}[1]{{\hat{\jmath}}_{#1}}

\renewcommand{\vec}[1]{\mathbf{#1}}

\newcommand{\commutator}[2]{\left[#1,#2\right]}
\newcommand{\ket}[1]{\left|#1\right>}
\newcommand{\bra}[1]{\left<#1\right|}
\renewcommand{\H}[0]{\hat{H}}
\newcommand{\dens}[1]{\hat{\sigma}_{#1}}

\newcommand{\mean}[1]{\left<#1\right>}

\newcommand{\var}[0]{\mathrm{Var}}

\newcommand{\vac}[1]{{\hat{V}}_{#1}}

\begin{document}

\begin{minipage}{\textwidth}
  \centering
  {\huge Deterministic atom-light quantum interface} \\
\vspace{0.5cm} 
  {\large Jacob Sherson$^{\#}$, Brian Julsgaard$^*$, 
           and E.~S.~Polzik}\\
\vspace{0.5cm} 
{\small   QUANTOP, Danish Research Foundation Center for Quantum Optics,\\
   Niels Bohr Institute, Blegdamsvej 17, DK-2100 Copenhagen \O, Denmark.\\
\vspace{0.1cm}
}
\vspace{0.5cm}
{\scriptsize
$^\#$ Other affilitation: 
Department of Physics and Astronomy, University of Aarhus,\\
   Ny Munkegade, bygning 520, DK-8000 Aarhus C, Denmark.\\
\vspace{0.1cm}
  $^*$ Present address: 
Lund Institute of Technology, Box 118, S-221 00 Lund, Sweden.\\
\vspace{0.1cm}
}
\end{minipage}

\hspace{0.05\linewidth}
\begin{minipage}[c]{0.9\linewidth}
  The notion of an atom-light quantum interface has been developed in
the past decade, to a large extent due to demands within the new
field of quantum information processing and communication. A
promising type of such interface using large atomic ensembles has
emerged in the past several years. In this article we review this
area of research with a special emphasis on deterministic high
fidelity quantum information protocols. Two recent experiments,
entanglement of distant atomic objects and quantum memory for
light are described in detail. 
\end{minipage}

\section{Introduction}

Interaction of light with atoms has been always one of the most
exciting areas in AMO physics. A complete quantum-mechanical approach
to it allows for considering such aspects as quantum state transfer
between atoms and light and generation of entangled states of light
and atoms. Entangled or non-separable states form the basis for
Quantum Information Processing and Communication where information is
encoded, processed and transferred in quantum states of light and
matter (for a collection of review articles on QIPC see, for example,
\citet{zoller:05}). Processing information encoded in quantum states
provides parallelism and unconditional security, two most promising
properties within this field. State transfer between matter where
quantum information is processed and stored, and light, the prime long
distance carrier of information, is crucial for quantum networks and
other applications.

A well-designed interaction between two or more quantum systems is an
efficient tool for creation of a desired joint entangled state of the
systems. Such interaction alone, or combined with a projective
measurement on one of the systems can lead to the creation of a
non-trivial quantum state of the remaining system(s). In this review
we describe recent progress in the development of a tool box for
deterministic quantum state generation of light and atomic ensembles
via a well known dipole interaction.  This tool box forms the basis
for a quantum atom-light interface capable of generating
entangled states of atoms and light on demand, performing efficient
quantum state exchange between light and atoms - quantum memory for
light, and long distance transfer of a quantum state of atoms via
teleportation and other quantum transfer protocols.

A key to successful quantum state engineering is a favorable balance
between efficient unitary interaction between the systems in question
(light and atoms within the context of this article) and decoherence
caused by interaction with the environment. Historically the first
setting considered most promising for the light-atoms interaction was
Cavity Quantum Electrodynamics (Cavity QED) where the Jaynes-Cummings
type of interaction could be realized. Driven by the interest in
single photon - single atom interaction~\citep{cirac:97}, this field
of research has achieved spectacular successes in both
optical~\citep{mckeever:04, kuhn:02} and microwave~\citep{raimond:01}
domains. Through the dramatic improvements in cavity quality factors
and with the use of cold atoms, coupling of single atoms to single
photon cavity modes has been demonstrated. However, formidable
technical challenges of the strong coupling cavity QED regime promoted
search for alternative routes to atom-light quantum interfaces.

One of the most successful of these emerged in the second half of the
90s with the recognition of the fact that the use of collective
quantum states of large atomic ensembles instead of single atoms can
provide efficient "strong" coupling to light outside of the cavity QED
regime. An intuitive explanation goes as follows. In order to be
efficiently coupled to a single atom, light has to be focused and
shaped to match the atomic dipole pattern.  Since this is very
difficult in free space, a cavity enhancing interaction of light with
a specific mode is necessary. On the other hand a collection of atoms
with transverse dimensions large compared to the wavelength of light
collectively couples very efficiently to any spatial light mode which
matches its shape. In classical electrodynamics this is manifested by
large absorption and dispersion that characterizes an atomic ensemble
interacting with nearly resonant light. Luckily, the same scaling
persists for some collective quantum properties of light and atoms.

The first approach to coupling quantum features of light to an atomic
ensemble was based on a straightforward assumption that if the input
light is completely absorbed by atoms, its quantum features should be
to a certain extent transferred onto atoms. The
proposal~\citep{kuzmich:97} considered a V-type atomic level scheme
where excitation in the two arms of the V-transition is carried out by
quantum correlated - entangled - light modes.  Absorption of these
modes leads to the creation of entanglement in the excited state of
atoms. If excitation, for example, is performed with squeezed light,
the excited state of atoms becomes spin squeezed. In case of cw
excitation spontaneous emission was shown to limit the degree of spin
squeezing of atoms to 50\%. The experimental implementation of this
proposal has been carried out for cold cesium atoms excited by
squeezed light resulting in the first demonstration of a macroscopic
($10^{6}$ atoms) ensemble of entangled atoms~\citep{hald:99}.

Efficient coupling at the quantum level also requires that no other
modes of light but the desired input one couple to the atoms. In
classical physics other modes are of no interest provided they do not
contain any light. In quantum physics any empty mode is in a vacuum
state. Coupling to such modes leads to spontaneous emission of atoms,
the major source of decoherence which a quantum state engineer has to
fight. In the Cavity QED setting interaction with vacuum modes of the
field is made small compared to coupling to the relevant input mode by
the strong coupling provided by the cavity.  How can this type of
decoherence be suppressed for a large atomic ensemble in the absence
of any cavity? The solution is to use for coupling with light two
substates of the atomic ground state rather than a ground and an
excited state as in the standard Jaynes-Cummings approach. The
advantages of using ground state levels are multiple. Their
spontaneous decay is negligible leading to long coherence times. Their
energy spacing is in the radio-frequency or microwave domain which
means that with the size of the ensemble smaller than the radio- or
microwave- wavelength the position of each atom is irrelevant and,
hence, collective coupling insensitive to atomic motion can be
achieved.

In order to circumvent limitations due to spontaneous emission, the idea of
complete absorption has been taken further by~\citet{kozhekin:00} where a
driven Raman transition involving a weak quantum mode in one arm has been
shown to be capable of faithful transfer of the light quantum state onto
\emph{ground state} collective atomic coherence. In a parallel development, a
celebrated electro-magnetically induced transparency process utilizing
carefully timed Raman pulses has been proposed for quantum storage of light
~\citep{fleischhauer:02}. The first experiment testing these ideas has been
recently carried out~\citep{eisaman:05}.

In all of the above approaches to light-atoms quantum interface the tool box
has only included light-atoms interactions. A significant next step allowing
implementation of atomic entanglement and quantum memory has been made by
adding a quantum measurement and feedback to the picture. The two experiments
which are central to this article, the deterministic entanglement of distant
atomic objects and the deterministic quantum memory for light, have been
carried out following essentially the same scenario. A pulse of light
interacts with atoms via a quantum non-demolition (QND) Hamiltonian, a
projective homodyning measurement is performed on the transmitted light, and
the feedback conditioned on the result of the measurement is applied to atoms
(Fig.~\ref{fig:twoCellSetup} and \ref{fig:exp_setup_typical}). Early proposals
for using the QND interaction for atomic quantum state engineering have been
published by~\citet{sanders:89},~\citet{corney:98}, and~\citet{wiseman:98}. In
particular, spin squeezing in an atomic ensemble generated by a QND
measurement has been studied theoretically in \citep{kuzmich:98,
takahashi:99, madsen:04} and experimentally demonstrated in
\citep{Kuzmichspinsq,geremia:04}. Quantum feedback has been explored in a
series of papers by Wiseman and co-authors. In particular quantum feedback in
relation to spin squeezing has been analyzed in~\citep{thomsen:02}.

To complete this brief introduction into atom-light quantum interface
research, we wish to draw a line between the deterministic interface reviewed
in this article and probabilistic schemes. The latter are based on generation
of entanglement between light and atoms conditioned on a random event of
detection of a photon in a certain mode. Spectacular experimental progress
along these lines has been achieved for schemes involving a single
ion~\citep{blinov:04,polzik:04}, single atom (\cite{weinfurter:lightatoment}),
and atomic ensembles~\citep{kuzmich:03, Wal:03, chaneliere:05, chou:05,
eisaman:05}.

The paper is organized as follows. In section 2 we introduce quantum variables
for light - Stokes operators, and for atoms - collective spin operators. An
effective Hamiltonian and equations of motion for the atomic and light
variables are then derived. We then expand the theory to two atomic ensembles
with an addition of a bias magnetic field.

In section 3 we formulate the equations of motion in the language of canonical
variables for light and atoms. Two main quantum information protocols are then
described theoretically: entanglement of two atomic ensembles and quantum
memory for light. In both case the treatment includes a simple decoherence
model. The quantum memory protocol which we dubbed "the direct mapping
protocol" represents an alternative to teleportation method of a high-fidelity
quantum state transfer from one system to another. Unlike teleportation it
involves only two objects: one carrying the input state and another the target
on which the input state is transferred.

Section 4 describes experimental methods and the key elements of the setup
including atomic cells, magnetic fields and detection of light. Application of
the magneto-optical resonance method for characterization of the collective
atomic spin states is outlined. Magnetic feedback used for the spin
manipulation is discussed.

Section 5 contains the main experimental results on atomic entanglement and
quantum memory for light. The determination of the benchmark atomic quantum
noise level - the projection noise - is described in detail. The effects
leading to decoherence of generated atomic entangled states are summarized.

In the appendices we discuss various technical aspects, such as the
effect of atomic motion on quantum state generation and decoherence,
the influence of polarization of light, and of technical noise of the
probe laser.

We conclude with a brief summary and outlook.

%
%
\section{Atom-Light Interaction}
\label{sec:light-atom-inter}

In this section we introduce the quantum variables for light and atoms
and describe the off-resonant dipole interaction between the
ground state $6S_{1/2}$ and the excited state $6P_{3/2}$ in
cesium. We will use spin operators for atoms and Stokes operators for
light as a convenient way do describe the interaction, and we
will present an effective Hamiltonian which describes the
dynamics of the ground state spin and the light. For simplicity the
Hamiltonian will be specific to the cesium ground state, although, the
same procedure can be applied for any other atom with a magnetically
non-degenerate ground state. With this as a starting point we derive
equations of motion for the light and atomic operators of a single
atomic ensemble.
  
\begin{figure}[t]
\centering
 \includegraphics[width=0.9\textwidth]{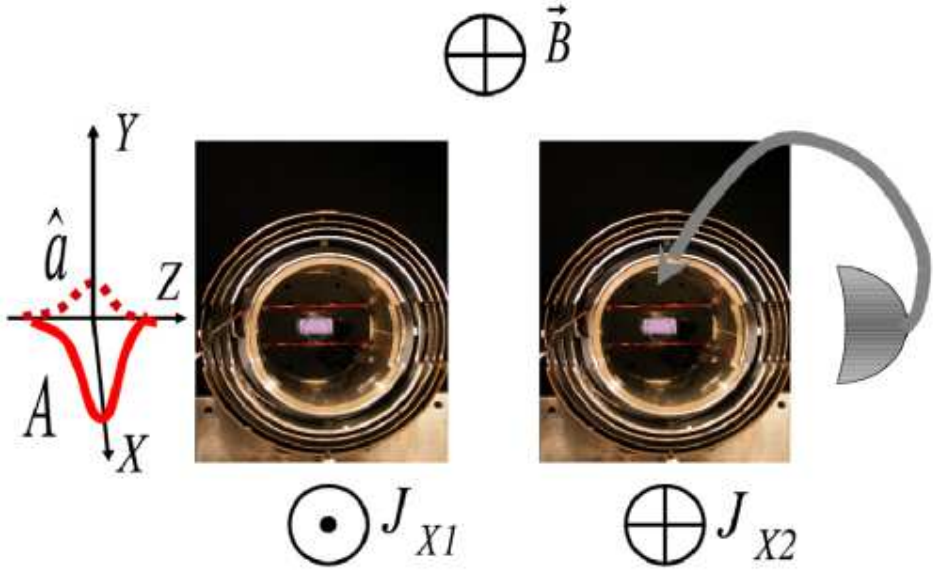}
\caption{\small 
  Schematic layout illustrating main components of the experiments.
  Two clouds of Cs atoms contained in paraffin coated glass cells are
  placed inside the magnetic shields. A light pulse consisting of
  orthogonally polarized strong classical and weak quantum fields
  passes through the cells and is detected by the detector on the
  right. Electronic feedback is applied to the rf magnetic coils
  surrounding cells to rotate the collective atomic spin. The bias
  magnetic field is applied to atoms in both cells. Atoms in the two
  cells are optically pumped as shown in the figure.  }
 \label{fig:twoCellSetup}
\end{figure}

A significant next step for quantum information protocols discussed in
this article is the transition from a single atomic ensemble to two
oppositely oriented ensembles accompanied by the addition of a
constant magnetic field (Fig.~\ref{fig:twoCellSetup}). If this field
is oriented along the mean atomic spin direction it allows for
achieving entanglement between two ensembles with a single pulse of
light instead of two pulses required in the absence of the field. An
even more important advantage is brought about by the possibility to
make measurements at a rather high Zeeman frequency, thus achieving
quantum limits of sensitivity with macroscopic numbers of atoms via
spectral filtering of classical noise.

\subsection{Atomic Spin Operators}
\label{sec:atomic_spin_operators}
The ground states of cesium are characterized by its outermost
electron which is in the $6S_{1/2}$ state, i.e.~the orbital angular
momentum $\vec{L}$ is zero. The electron spin $\vec{S}$ and thus the
total electronic angular momentum $\vec{J}$ has quantum number
$S=J=1/2$. The nuclear spin $\vec{I}$ of cesium-133 has $I=7/2$, and
the coupling between the nucleus and the electron gives rise to the
total angular momentum $\vec{F} = \vec{I} + \vec{J}$ with quantum
numbers $F=3$ and $F=4$. 

It is indeed the total angular momentum $\vec{F}$ which interests us
in this work since $F$ and the magnetic quantum numbers $m_F$ define
the energy levels of the ground states in the limit of low magnetic
field discussed here. Furthermore, we restrict ourselves to one
hyperfine level, $F=4$, which is possible experimentally since the
hyperfine splitting $\nu_{\mathrm{hfs}} = 9.1926\mathrm{GHz}$ is large
compared to typical resolutions of our laser systems. We choose to
denote the total angular momentum of a \emph{single atom} by $\vec{j}$
and for a collection of atoms (in the $F=4$ state) we denote the
\emph{collective} total angular momentum by $\vec{J}$, i.e. 
\begin{equation}
  \label{eq:define_J_collective}
  \vec{J} = \sum_{i = 1} ^ {N} \vec{j}^{(i)},
\end{equation}
where $N$ is the number of atoms in the $F=4$ state and
$\vec{j}^{(i)}$ is the total angular momentum of the $i$'th atom. The
reason for using $\vec{J}$ and not $\vec{F}$ is conventional, and
indeed, we wish to think about our spins more abstractly than just the
properties of some atoms. Many of the results should be applicable in
a broader sense than to a collection of cesium atoms. 

In our experiments the number of atoms $N$ is of order
$10^{12}$ and we will almost always aim at having all atoms polarized
along one direction which we denote as the $x$-axis. With the $x$-axis
as quantization axis we have $m_F = 4$ for all atoms to a high
degree of accuracy, and the collective spin $\J{x}$ will really be a
macroscopic entity. With this experimental choice, we may treat the
$x$-component of the collective spin as a classical $c$-number,
i.e.~we replace the operator $\J{x}$ by the number $J_x$. The
transverse spin components $\J{y}$ and $\J{z}$ maintain their quantum
nature. They will typically have zero or a small mean value. The
quantum fluctuations are governed by the commutation relation and the
Heisenberg uncertainty relation (with $\hbar = 1$)
\begin{gather}
\label{comm:JyJz}
  \commutator{\J{y}}{\J{z}} = i J_x \\
\label{Heisenberg:JyJz}  
  \Rightarrow\quad  \var(\J{y})\cdot\var(\J{z}) \ge \frac{J_x^2}{4}. 
\end{gather}
With $10^{12}$ atoms the quantum uncertainty of the angle of the
collective spin direction is of order $10^{-6}$.

\subsection{Polarization States of Light}
\label{sec:pol_states_light}
All our experiments involve narrow-band light interacting with atomic
spin states, and it turns out that the polarization states of the
light form a convenient language to describe the light degrees of
freedom. 

Consider a pulse of light, or a collection of photons,
propagating in the $z$-direction. The polarization state is well
described by the Stokes operators
\begin{equation}
  \label{def:intro_stokes}
  \begin{split}
    \S{x} &= \frac{1}{2}\left(\hat{n}_{\mathrm{ph}}(x) - 
                               \hat{n}_{\mathrm{ph}}(y)\right)
           = \frac{1}{2}\left(\adag{x}\a{x} - \adag{y}\a{y}\right), \\
    \S{y} &= \frac{1}{2}\left(\hat{n}_{\mathrm{ph}}(+45^{\circ}) -
                              \hat{n}_{\mathrm{ph}}(-45^{\circ})\right)
           = \frac{1}{2}\left(\adag{x}\a{y} + \adag{y}\a{x}\right), \\
    \S{z} &= \frac{1}{2}\left(\hat{n}_{\mathrm{ph}}(\sigma_+) -
                              \hat{n}_{\mathrm{ph}}(\sigma_-)\right)
           = \frac{1}{2i}\left(\adag{x}\a{y} - \adag{y}\a{x}\right),
  \end{split}
\end{equation}
where $\hat{n}_{\mathrm{ph}}(x)$ is the number of photons in the pulse
with $x$-polarization, and so on. The Stokes operators are
dimensionless as written here, they count photons. At our convenience
we will later break these up into time or spatial slices. 

We make the following assumption central to the experiments
described below. We assume that light consists of a strong component
linearly polarized along the $x$-direction and a much weaker component
polarized in the $y$-plane (Fig.~\ref{fig:twoCellSetup}). This means
that we can treat the $x$-mode operators $\a{x}$, $\adag{x}$, and
hence $\S{x} \rightarrow S_x$ as a $c$-number.  Note, this is very
similar to the approximation of a well polarized sample of spins in
the previous section.  Specifically we find (with $\adag{x} = \a{x} =
A_x$ being real) that $\S{y} = A_x\cdot(\a{y}+\adag{y})/2$ and $\S{z}
= A_x\cdot(\a{y}-\adag{y})/2i$. We see that in our approximation the
quantum properties of $\S{y}$ and $\S{z}$ are solely encoded in the
$y$-polarized part of the light.

It can be shown that the Stokes vector satisfies angular commutation
relations
\begin{gather}
\label{comm:SySz_intro}
  \commutator{\S{y}}{\S{z}} = i S_x \\
\label{Heisenberg:SySz}  
  \Rightarrow\quad  \var(\S{y})\cdot\var(\S{z}) \ge \frac{S_x^2}{4}
  \quad\text{(pulse of light)}. 
\end{gather}

\subsection{Off-Resonant Coupling}
\label{sec:off-resonant-coupling}
We consider the special case with a propagating beam of light
coupled off-resonantly to the $6S_{1/2,F=4}\rightarrow
6P_{3/2,F'=3,4,5}$ transitions in cesium. Then absorption effects can
be neglected and all the dynamics are of dispersive nature.
Furthermore, the optically excited states can be adiabatically
eliminated and an effective Hamiltonian describing the light
interacting with only ground state degrees of freedom is obtained
\citep{julsgaard_phd,
  hammerer:teleportation}\footnote{\citep{julsgaard_phd} contains a
  factor of two error which has been corrected in
  Eq.~(\ref{sec_inter:final_eff_Hamil_real}).}.
%
%
%
%

\begin{equation}
\begin{split}
\label{sec_inter:final_eff_Hamil_real}
\H_{\mathrm{int}}^{\mathrm{eff}} = &-\frac{\hbar c\gamma}{8 A \Delta}
\frac{\lambda^2}{2\pi}\int_0^L \left(
a_0\cdot\hat{\phi}(z,t) + a_1\cdot \S{z}(z,t)\j{z}(z,t) \right.\\
&+ \left. a_2\left[
\hat{\phi}(z,t)\j{z}^2(z,t) - \S{-}(z,t)\j{+}^2(z,t) - \S{+}(z,t)\j{-}^2(z,t) 
\right] \right) \rho A dz. 
\end{split}
\end{equation}
Here we have assumed a one-dimensional theory for the light which is
sufficient for a beam cross section $A$ that is much larger than the
squared wavelength $\lambda^2$. The small letter spin operators
$\j{}(z,t)$ are dimensionless and refer to single atoms at position
$z$ at time $t$. The Stokes operators $\S{}(z,t)$ are taken to count
number of photons per unit length at position $z$ and time $t$. The
integration then runs over the entire sample of length $L$ with atomic
density $\rho$. In the front factor $\gamma$ is the natural FWHM line
width of the optical transition $6S_{1/2}\rightarrow 6P_{3/2}$ and
$\Delta$ is the detuning from the $F=4$ to $F'=5$ transition with red
being positive. 

As for the operators, $\hat{\phi}(z,t)$ is the photon flux per length,
$\S{+} = \S{x}+i\S{y} = -\adag{+}\a{-}$ and $\S{-} = \S{x}-i\S{y} =
-\adag{-}\a{+}$ are raising and lowering operators converting
$\sigma_+$-photons into $\sigma_-$-photons or vice versa, $\j{\pm} =
\j{x}\pm i\j{y}$ are the usual raising and lowering operators for the
spin. 

The parameters $a_0$, $a_1$, and $a_2$ are for the $F=4$ ground state
in cesium given by
\begin{equation}
\label{def:a0a1a2_F=4}
\begin{split}
a_0 &= \frac{1}{4}
\left(\frac{1}{1-\Delta_{35}/\Delta}+\frac{7}{1-\Delta_{45}/\Delta}
    + 8\right) \rightarrow 4, \qquad\qquad(F=4)\\
a_1 &= \frac{1}{120}
\left(-\frac{35}{1-\Delta_{35}/\Delta}-\frac{21}{1-\Delta_{45}/\Delta}
    + 176\right) \rightarrow 1,\\
a_2 &= \frac{1}{240}
\left(\frac{5}{1-\Delta_{35}/\Delta}-\frac{21}{1-\Delta_{45}/\Delta}
    + 16\right) \rightarrow 0, 
\end{split}
\end{equation}
where the limit is calculated for very large values of the detuning.
The detunings $\Delta _{35}/2\pi =452.2$ MHz and $\Delta _{45}/2\pi
=251.0$ MHz are given by the splitting in the excited state. Let us
comment on the different terms in the
Hamiltonian~(\ref{sec_inter:final_eff_Hamil_real}). The first term
containing $a_0$ just gives a Stark shift to all atoms independent of
the internal state but proportional to the photon density
$\hat{\phi}(z,t)$. The second term containing $a_1$ rotates the Stokes
vector $\vec{S}$ and the spin vector $\vec{J}$ around the $z$-axis,
known as Faraday rotation - we look more closely into this below. The
last terms proportional to $a_2$ are higher order couplings between
the light and the atoms and since $a_2$ is small for a sufficiently
large detuning these can normally be neglected.

All these terms conserve individually the $z$-projection of the total
angular momentum of light and atoms, e.g.~the $\S{-}\j{+}^2$ term can
change a $\sigma_+$ photon into a $\sigma_-$ photon (changing the
light angular momentum along $z$ by $-2\hbar$ while the atoms receive
$2\hbar$ mediated by the atomic raising operator $\j{+}^2$. The total
angular momentum must have its $z$-projection invariant since the
physical system is axially symmetric around the direction of light
propagation (the $z$-axis). 

For us the term proportional to $a_1$ is useful and relevant.  This
term represents the QND interaction. The higher order terms
proportional to $a_2$ create some minor problems which can be
minimized with large detuning (this will be discussed further in
sections~\ref{sec:polarization_and_Stark} and~\ref{sec:laser_noise}). 
For a detailed treatment of the higher order - atomic alignment -
effects we refer to~\citep{kupriyanov:05}.  The zeroth order term
proportional to $a_0$ produces an overall phase shift and can be
omitted. 

\subsection{Propagation Equations}
The Hamiltonian~(\ref{sec_inter:final_eff_Hamil_real}) is a very
convenient starting point for many calculations and we now show
the procedure to derive equations of motion. For the spin operators we
need the Heisenberg evolution $\partial \j{i}/\partial t =
\frac{1}{i\hbar}\commutator{\j{i}}{\H}$ and for the Stokes operators
the Maxwell-Bloch equations $(\partial/\partial t + c\cdot\partial
/\partial z)\S{i}(z,t) =
\frac{1}{i\hbar}\commutator{\S{i}(z,t)}{\H_{\mathrm{int}}}$, see
\citep{julsgaard_phd}. For the latter we will neglect retardation
effects, i.e.~we do not calculate dynamics on the time scale $L/c$ of
propagation across the sample. This is equivalent to setting the speed
of light to infinity and hence leaving out the $\partial/\partial t$
term.  If we consider only the term proportional to $a_1$ and neglect
the other we find
\begin{equation}
\label{eq:propagation_spin}
  \begin{split}
  \frac{\partial \j{x}(z,t)}{\partial t} 
    &= +\frac{c\gamma}{8A\Delta}\frac{\lambda^2}{2\pi}a_1 \S{z}(z,t)j_y(z,t),\\
  \frac{\partial \j{y}(z,t)}{\partial t} 
    &= -\frac{c\gamma}{8A\Delta}\frac{\lambda^2}{2\pi}a_1 \S{z}(z,t)j_x(z,t),\\
  \frac{\partial \j{z}(z,t)}{\partial t} &= 0. 
  \end{split}
\end{equation}
and
\begin{equation}
\label{eq:propagation_light}
  \begin{split}
  \frac{\partial}{\partial z} \S{x}(z,t) &=
   +\frac{\gamma\rho}{8\Delta}\frac{\lambda^2}{2\pi}
    a_1\S{y}(z,t)\j{z}(z,t), \\
  \frac{\partial}{\partial z} \S{y}(z,t) &=
   -\frac{\gamma\rho}{8\Delta}\frac{\lambda^2}{2\pi}
    a_1\S{x}(z,t)\j{z}(z,t), \\
  \frac{\partial}{\partial z} \S{z}(z,t) &= 0. 
  \end{split}
\end{equation}
We observe that $\j{z}(z,t)$ and $\S{z}(z,t)$ are individually
conserved during the interaction which is also apparent from the
Hamiltonian~(\ref{sec_inter:final_eff_Hamil_real}) since these
operators commute with the $a_1$-term. The effect of the interaction
is that the spin will rotate around the $z$-axis with an amount
proportional to $\S{z}$, and the Stokes vector will rotate around the
$z$-axis proportionally with $\j{z}$. 

Below we assume that these rotations are small and that the
dominant classical (mean) polarization vector of light and the
orientation vector of the collective atomic spin stay oriented along
the $x$-direction after the interaction. This turns out to be a very
good approximation for experimentally attainable values of the
interaction strength. Under this assumption the first line of the
systems ~(\ref{eq:propagation_spin}) and~(\ref{eq:propagation_light}),
respectively, can be omitted.  Furthermore, given the QND structure of
the remaining equations, we can sum over the individual atomic spins
and obtain the same equation for the collective spin
variables~(\ref{eq:define_J_collective}). In our continuous notation
we have $\J{i}(t) = \int_0^L \j{i}(z,t)\rho A dz$. As for the
light operators we concentrate on the in- and out-going parts
only. Hence we define $\S{i}^{\mathrm{in}} = c\S{i}(z=0,t)$ and
$\S{i}^{\mathrm{out}} = c\S{i}(z=L,t)$. The multiplication by the
speed of light $c$ turns the normalization into photons per unit time
instead of per unit length. With the assumption of small rotation angles,
integrating Eqs.~(\ref{eq:propagation_spin})
and~(\ref{eq:propagation_light}) over space from $z=0$ to $z=L$ leads
to the following very important equations:
\begin{align}
\label{eq:syout}
  \S{y}^{\mathrm{out}}(t) &= \S{y}^{\mathrm{in}}(t) + a S_x \hat{J}_z(t), \\
\label{eq:szout}
  \S{z}^{\mathrm{out}}(t) &= \S{z}^{\mathrm{in}}(t), \\
\label{eq:jyout}
  \frac{\partial}{\partial t}\J{y}(t) &= a J_x \S{z}^{\mathrm{in}}(t), \\
\label{eq:jzout}
  \frac{\partial}{\partial t}\J{z}(t) &= 0,
\end{align}
where $a = - \frac{\gamma }{8 A \Delta}\frac{\lambda^2}{2\pi}a_1$.  In
and out refer to light before and after passing the atomic sample,
respectively.

We note from Eqs.~(\ref{eq:syout}) and (\ref{eq:jzout}) that in the
case of a large interaction strength (i.e.~if $aS_x\J{z}$ dominates
$\S{y}^{\mathrm{in}}$) a measurement on $\S{y}^{\mathrm{out}}$ amounts
to a QND measurement of $\J{z}$.  Using off-resonant light for QND
measurements of spins has also been discussed in \citep{kuzmich:98,
  takahashi:99}. Equation~(\ref{eq:jyout}) implies that a part of the state
of light is also mapped onto the atoms - we denote this as back
action. This opens up the possibility of using this sort of system for
quantum memory which will be the topic of
sections~\ref{sec:quantum-memory-protocols}
and~\ref{sec:quant-map-results}.

\subsection{The Rotating Frame}
\label{sec:addbfield}
In the experiment a constant and homogeneous magnetic field is added
in the $x$-direction. We discuss the experimental reason for this
below. For our modeling, the magnetic field adds a term $H_B=\Omega
J_x$ to the Hamiltonian. This makes the transverse spin components
precess at the Larmor frequency $\Omega$ depending on the strength of
the field.  Introducing the rotating frame coordinates:
\begin{equation}
\label{eq:rotating_frame_dynamics}
\left(
\begin{array}{c}
  \J{y}' \\
  \J{z}' \\
\end{array}
\right)
=
\left(
\begin{array}{cc}  \cos(\Omega t) & \sin(\Omega t) \\
  -\sin(\Omega t) & \cos(\Omega t) \\
\end{array}
\right)
\left(
\begin{array}{c}
  \J{y} \\
  \J{z} \\
\end{array}
\right)
\end{equation}
we can easily show that Eqs.~(\ref{eq:syout})-(\ref{eq:jzout})
transform into:
\begin{align}
\label{eq:Syout_rotating}
  \S{y}^{\mathrm{out}}(t) &= \S{y}^{\mathrm{in}}(t) + 
    a S_x\left(\J{y}'(t)\sin(\Omega t) + \J{z}'(t)\cos(\Omega t)\right), \\
\label{eq:Szout_rotating}
  \S{z}^{\mathrm{out}}(t) &= \S{z}^{\mathrm{in}}(t), \\
\label{eq:Jydot_rotating}
  \frac{\partial}{\partial t} \J{y}'(t)
    &= a J_x \S{z}^{\mathrm{in}}(t)\cos(\Omega t), \\
\label{eq:Jzdot_rotating}
  \frac{\partial}{\partial t} \J{z}'(t) &= a J_x
    \S{z}^{\mathrm{in}}(t)\sin(\Omega t). 
\end{align}
Thus, the atomic imprint on the light is encoded in the
$\Omega$-sideband instead of at the carrier frequency. The primary
motivation for adding the magnetic field is the fact that lasers are
generally a lot more quiet at high sideband frequencies compared to
the carrier. A measurement without a magnetic field will be a DC
measurement and the technical noise would dominate the subtle quantum
signal. Also, as the measurement time is longer than $1/\Omega$
Eq.~(\ref{eq:Syout_rotating}) enables us to access both $J_{y}'$ and
$J_{z}'$ at the same time. We are of course not allowed to perform
non-destructive measurements on these two operators simultaneously
since they are non-commuting. This is also reflected by the fact that
neither $\J{y}$ nor $\J{z}$ are constant in
Eqs.~(\ref{eq:Jydot_rotating}) and~(\ref{eq:Jzdot_rotating}). Below we
shall consider two atomic samples and we will see that a QND-type
interaction can be regained in this setting.

\subsection{Two Oppositely Oriented Spin Samples}
Inspired by the above we now assume that we have two atomic
samples with oriented spins such that $J_{x1} = -J_{x2} \equiv J_x$. 
We re-express the equations of
motion~(\ref{eq:Syout_rotating})-(\ref{eq:Jzdot_rotating}) for two
samples in a way which is much more convenient for the understanding
of our entanglement creation and verification procedure. 

For two atomic samples we write equations of motion:
\begin{align}
\label{eq:Syout_rotating_two}
  \S{y}^{\mathrm{out}}(t) &= \S{y}^{\mathrm{in}}(t) + 
    a S_x\left([\J{y1}'(t)+\J{y2}'(t)]\sin(\Omega t)\right. \\ \notag
      &\left.\qquad\qquad\qquad\;
       + [\J{z1}'(t)+\J{z2}'(t)]\cos(\Omega t)\right), \\
\label{eq:Jy1y2dot_rotating}
  \frac{\partial}{\partial t} (\J{y1}'(t)+\J{y2}'(t))
    &= a (J_{x1}+J_{x2})\S{z}^{\mathrm{in}}(t)\cos(\Omega t) = 0, \\
\label{eq:Jz1z2dot_rotating}
  \frac{\partial}{\partial t} (\J{z1}'(t)+\J{z2}'(t)) 
    &= a (J_{x1}+J_{x2})\S{z}^{\mathrm{in}}(t)\sin(\Omega t) = 0. 
\end{align}
The fact that the sums $\J{y1}'(t)+\J{y2}'(t)$ and
$\J{z1}'(t)+\J{z2}'(t)$ have zero time derivative relies on the
assumption of opposite spins of equal magnitude. The constancy of
these terms together with Eq.~(\ref{eq:Syout_rotating_two}) allows us
to perform QND measurements on the two sums. We note that each of the
sums can be accessed by considering the two operators
\begin{align}
  \int_0^T \S{y}^{\mathrm{out}}\cos(\Omega t)dt &= 
    \int_0^T \S{y}^{\mathrm{in}}\cos(\Omega t)dt 
    + \frac{a S_x T}{2}(\J{z1}'(t)+\J{z2}'(t)), \\
  \int_0^T \S{y}^{\mathrm{out}}\sin(\Omega t)dt &= 
    \int_0^T \S{y}^{\mathrm{in}}\sin(\Omega t)dt 
    + \frac{a S_x T}{2}(\J{y1}'(t)+\J{y2}'(t)). 
\end{align}
We have used the fact that $\int_0^T\cos^2(\Omega t)dt \approx
\int_0^T\sin^2(\Omega t)dt \approx T/2$ and that $\int_0^T\cos(\Omega
t)\sin(\Omega t)dt \approx 0$. Each of the operators on the left hand
side can be measured simultaneously by making a $\S{y}$-measurement
and multiplying the photo-current by $\cos(\Omega t)$ or $\sin(\Omega
t)$ followed by integration over the duration $T$. The possibility to
gain information about $\J{y1}'(t)+\J{y2}'(t)$ and
$\J{z1}'(t)+\J{z2}'(t)$ enables us to generate entangled states, the
topic of sections~\ref{sec:ent_gen_ver}
and~\ref{sec:entanglement-results}. At the same time we must lose
information about some other physical variable. This is indeed true,
the conjugate variables to these sums are $\J{z2}'(t)-\J{z1}'(t)$ and
$\J{y1}'(t)-\J{y2}'(t)$, respectively.  These have the time evolution
\begin{align}
  \frac{\partial}{\partial t} (\J{y1}'(t)-\J{y2}'(t))
    &= 2a J_x\S{z}^{\mathrm{in}}(t)\cos(\Omega t), \\
  \frac{\partial}{\partial t} (\J{z1}'(t)-\J{z2}'(t)) 
    &= 2a J_x\S{z}^{\mathrm{in}}(t)\sin(\Omega t).  
\end{align}
We see how noise from the input $\S{z}$-variable is piling up in the
difference components while we are allowed to learn about the sum
components via $\S{y}$ measurements. The above equations clearly
describe the physical ingredients for light and atoms.  However, when
transfer of a quantum state between two very different systems - light
and atoms - is concerned, an isomorphic formulation, such as rendered
by canonical variables, is desirable.  In Sec.~\ref{sec:protocols} we
will re-write the theory in this more convenient language.

%
%
\section{Quantum Information Protocols}
\label{sec:protocols}
In this chapter we introduce in detail two main protocols of this
paper: deterministic generation of entanglement of two macroscopic
objects and deterministic quantum state mapping from one system onto
another - the so-called direct mapping protocol.  The latter is then
applied as the quantum memory for light protocol.

In order to simplify the description of the protocols we re-write
equations of motion of the previous section in the language of
canonical variables. This also allows for unified treatment of a
single ensemble without- and two ensembles, with a constant magnetic
field. For a single sample we define:
\begin{equation}
\begin{array}{ll}
  \X{\mathrm{A}s} = \frac{\J{y}}{\sqrt{J_x}},&
  \P{\mathrm{A}s} = \frac{\J{z}}{\sqrt{J_x}},  \\
  \X{\mathrm{L}s} = \frac{1}{\sqrt{S_x T}}\int_0^T\S{y}(t)dt, &
  \P{\mathrm{L}s} = \frac{1}{\sqrt{S_x T}}\int_0^T\S{z}(t)dt,
\end{array}
\end{equation}
For two samples we get two sets of canonical variables by defining the
atomic operators:
\begin{subequations}
\begin{align}
  \X{\mathrm{A}1} &= \frac{\J{y1}'-\J{y2}'}{\sqrt{2J_x}},  \\
  \P{\mathrm{A}1} &= \frac{\J{z1}'+\J{z2}'}{\sqrt{2J_x}},  \\
  \X{\mathrm{A}2} &= -\frac{\J{z1}'-\J{z2}'}{\sqrt{2J_x}}, \\
  \P{\mathrm{A}2} &= \frac{\J{y1}'+\J{y2}'}{\sqrt{2J_x}}.
\end{align}
\end{subequations}
and the light operators:
\begin{subequations}
\begin{align}
\label{eq:def_XL1}
  \X{\mathrm{L}1} &= \sqrt{\frac{2}{S_x T}}\int_0^T\S{y}(t)\cos(\Omega t)dt, \\
  \P{\mathrm{L}1} &= \sqrt{\frac{2}{S_x T}}\int_0^T\S{z}(t)\cos(\Omega t)dt, \\
  \X{\mathrm{L}2} &= \sqrt{\frac{2}{S_x T}}\int_0^T\S{y}(t)\sin(\Omega t)dt, \\
  \P{\mathrm{L}2} &= \sqrt{\frac{2}{S_x T}}\int_0^T\S{z}(t)\sin(\Omega t)dt.
\end{align}
\end{subequations}
Each pair of $\X{},\P{}$ operators satisfy the usual commutation
relation, e.g.~we have $\commutator{\X{\mathrm{L}1}}{\P{\mathrm{L}1}}
= i$. Equations (\ref{eq:syout}-\ref{eq:jzout}) and
(\ref{eq:Syout_rotating_two}-\ref{eq:Jz1z2dot_rotating}) now translate
into
\begin{subequations}
\begin{align}
\label{interact_light}
  \X{\mathrm{L}i}^{\mathrm{out}} &= \X{\mathrm{L}i}^{\mathrm{in}} +
    \kappa\P{\mathrm{A}i}^{\mathrm{in}},\\
  \P{\mathrm{L}i}^{\mathrm{out}} &= \P{\mathrm{L}i}^{\mathrm{in}},\\
\label{interact_atom}
  \X{\mathrm{A}i}^{\mathrm{out}} &= \X{\mathrm{A}i}^{\mathrm{in}} +
    \kappa\P{\mathrm{L}i}^{\mathrm{in}},\\
  \P{\mathrm{A}i}^{\mathrm{out}} &= \P{\mathrm{A}i}^{\mathrm{in}},
\end{align}
\label{eq:main_equations}
\end{subequations}
where we recall $i=1,2,s$ refer to the definitions above and not the
two samples. Note, that in the case of two samples the two sets of
interacting light and atomic operators are decoupled. The parameter
describing the strength of light/matter-interactions is given by
$\kappa = a\sqrt{J_x S_x T}$.  The limit to strong coupling is around
$\kappa \approx 1$. This set of equations represents the starting
point for numerous applications, which in the context of this paper
are implemented in the double sample setup.

\subsection{Entanglement - Two Mode Squeezing Protocol}
\label{sec:ent_gen_ver}
It follows from Eqs. (\ref{interact_light}) and (\ref{interact_atom})
that a sufficiently precise measurement of $\X{\mathrm{L}}$ of a light
pulse transmitted through two atomic ensembles renders knowledge about
$\P{\mathrm{A}}$. The measurement will project $\P{\mathrm{A}}$ into a
two-mode squeezed or Einstein-Podolsky-Rosen entangled
state~\citep{ourNature04}.

The necessary and sufficient condition for such an entangled state has
been derived by \citet{duanentcrit}. Demonstration of entanglement is
thus reduced to the fulfillment of this criterion. The criterion,
which is introduced shortly, utilizes the vacuum state noise of a
canonical variable, which for atomic ensembles reduces to the coherent
spin state fluctuations.  Hence, these fluctuations, which are also
called projection noise, form an extremely important benchmark in
experiments with atomic ensembles.
\subsubsection{The Coherent Spin State}
To create entangled or squeezed states one has to generate states
which exhibit less fluctuations than all equivalent classical states.
The boundary occurs at the \emph{coherent spin state} (CSS) in which all
spins are independent realizations of a single spin oriented along a
specific direction. The characteristics of this state are
discussed in more detail in Sec. \ref{sec:proj-noise-level} so for now
we need only to quantify the role of the coherent spin state as a
boundary between classical and purely quantum mechanical states. From
the canonical atomic operators a Heisenberg uncertainty relation can
be formed:
\begin{equation}
  \label{eq:heisenberg}
  \var(\X{\mathrm{A}i})\var(\P{\mathrm{A}i}) \ge
    \frac{\left|\mean{[\X{\mathrm{A}i},\P{\mathrm{A}i}]}\right|^2}{4}
    =\frac{1}{4}.
\end{equation}
For the coherent spin state the two variances are both equal to one
half, thus confirming that it is the classical state with the least
possible noise. For a state to be squeezed, it has to have less noise
in one of the quadratures. Since the Heisenberg uncertainty relation
still has to be fulfilled it follows that the other quadrature has to
exhibit excessive fluctuations, it is anti-squeezed. Entanglement is
the non-local interconnection of two systems, such that it is
impossible to write the total density matrix as a product of density
matrices for each system. It has been shown that for effectively
continuous variable systems such as the ones described so far in this
paper the necessary and sufficient criterion for entanglement
is~\citep{duanentcrit}
\begin{equation}
\label{eq:ent_critetrion_xp}
  \var(\P{\mathrm{A}1}) + \var(\P{\mathrm{A}2}) < 1.
\end{equation}
\subsubsection{Entanglement Generation and Verification}
We now turn to the actual understanding of entanglement generation and
verification. We concentrate here on generation of an entangled state
conditioned on the result of a measurement. Generation of
an unconditional entangled state with the help of feedback is described
in the experimental section below.  For generation of conditional
entanglement we perform the following steps: First the atoms are
prepared in the oppositely oriented coherent states corresponding to
creating the vacuum states of the two modes
$(\X{\mathrm{A}1},\P{\mathrm{A}1})$ and
$(\X{\mathrm{A}2},\P{\mathrm{A}2})$. Next a pulse of light called the
\emph{entangling pulse} is sent through the atoms and we measure the
two operators $\X{\mathrm{L}1}^{\mathrm{out}}$ and
$\X{\mathrm{L}2}^{\mathrm{out}}$ with outcomes $A_1$ and $B_1$,
respectively.  These results bear information about the atomic
operators $\P{\mathrm{A}1}$ and $\P{\mathrm{A}2}$ and hence we reduce
variances $\var(\P{\mathrm{A}1})$ and $\var(\P{\mathrm{A}2})$. To
prove we have an entangled state we must confirm that the variances of
$\P{\mathrm{A}1}$ and $\P{\mathrm{A}2}$ fulfill the
criterion~(\ref{eq:ent_critetrion_xp}). That is we need to know the
mean values of $\P{\mathrm{A}1}$ and $\P{\mathrm{A}2}$ with a total
precision better than unity. To demonstrate that, we send a second
\emph{verifying pulse} through the atomic samples again measuring
$\X{\mathrm{L}1}^{\mathrm{out}}$ and $\X{\mathrm{L}2}^{\mathrm{out}}$
with outcomes $A_2$ and $B_2$. Now it is a matter of comparing $A_1$
with $A_2$ and $B_1$ with $B_2$. If the results are sufficiently close
the state created by the first pulse was entangled.

Let us be more quantitative. The interaction~(\ref{interact_light})
mapping the atomic operators $\P{\mathrm{A}i}$ to field operators
$\X{\mathrm{L}i}$ is very useful for large $\kappa$ and useless if
$\kappa \ll 1$. We will describe in detail the role of $\kappa$ for
all values. To this end we first describe the natural way to determine
$\kappa$ experimentally. If we repeatedly perform the first two steps
of the measurement cycle, i.e.~prepare coherent states of the atomic
spins, send in the first measurement pulse, and record outcomes $A_1$
and $B_1$, we may deduce the statistical properties of the
measurements. Theoretically we expect from~(\ref{interact_light})
\begin{equation}
\label{eq:stat_projection}
  \mean{A_1} = \mean{B_1} = 0
  \quad\text{and}\quad
  \var(A_1) = \var(B_1) = \frac{1}{2} + \frac{\kappa^2}{2}.
\end{equation}
The first term in the variances is the shot noise (SN) of light.  This
can be measured in the absence of the interaction where $\kappa = 0$.
The quantum nature of the shot noise level is confirmed by checking
the linear scaling with photon number of the pulse. The second term
arises from the projection noise (PN) of atoms. Hence, we may
calibrate $\kappa^2$ to be the ratio $\kappa^2 = \mathrm{PN/SN}$ of
atomic projection noise to shot noise of light. Theoretically
$\kappa^2$ has linear scaling $\kappa^2 = a J_x S_x T$ with a 
macroscopic spin $J_x$ that must be confirmed in the experiment
(see Sec.~\ref{sec:proj-noise-level}).

Next we describe how to deduce the statistical properties of the state
created by the $\emph{entangling pulse}$. Based on the measurement
results $A_1$ and $B_1$ of this pulse we must predict the mean value
of the second measurement outcome. If $\kappa \rightarrow \infty$ we
ought to trust the first measurement completely since the initial
noise of $\X{\mathrm{L}i}^{\mathrm{in}}$ is negligible,
i.e.~$\mean{A_2} = A_1$ and $\mean{B_2} = B_1$. On the other hand, if
$\kappa = 0$ we know that atoms must still be in the vacuum state such
that $\mean{A_2} = \mean{B_2} = 0$. It is natural to take in general
$\mean{A_2} = \alpha A_1$ and $\mean{B_2} = \alpha B_1$. We need not
know a theoretical value for $\alpha$. The actual value can be deduced
from the data. If we repeat the measurement cycle $N$ times with
outcomes $A_1^{(i)}$, $B_1^{(i)}$, $A_2^{(i)}$, and $B_2^{(i)}$, the
correct $\alpha$ is found by minimizing the conditional variance
\begin{equation}
\label{eq:alpha}
\begin{split}
  \var(A_2|A_1) + \var(B_2|B_1) &= \\
  \min_{\alpha} \frac{1}{N-1}\sum_i^N &\left(
    (A_2^{(i)} - \alpha A_1^{(i)})^2 + (B_2^{(i)} - \alpha B_1^{(i)})^2
    \right).
\end{split}
\end{equation}
In order to deduce whether we fulfill the entanglement
criterion~(\ref{eq:ent_critetrion_xp}) we compare the above to our
expectation from~(\ref{interact_light}). For the verifying pulse we
get
\begin{equation}
  \begin{split}
   \mean{\left(\X{\mathrm{L}i}^{\mathrm{out}} -
   \mean{\X{\mathrm{L}i}^{\mathrm{out}}}\right)^2}
     &= \mean{\left(\X{\mathrm{L}i}^{\mathrm{in,2nd}} + \kappa\left[
     \P{\mathrm{A}i}^{\mathrm{ent}} -
     \mean{\P{\mathrm{A}i}^{\mathrm{ent}}}\right]\right)^2}\\
     &= \frac{1}{2} + \kappa^2 \var(\P{\mathrm{A}i}^{\mathrm{ent}}),
  \end{split}
\end{equation}
where $\X{\mathrm{L}i}^{\mathrm{in,2nd}}$ refers to the incoming light
of the \emph{verifying pulse} which has zero mean.
$\P{\mathrm{A}i}^{\mathrm{ent}}$ refers to the atoms after being
entangled.  We see that the practical entanglement criterion becomes
\begin{equation}
\label{eq:easy_critertion}
  \begin{split}
      \var(A_2|A_1) + \var(B_2|B_1)
      &= 1 + \kappa^2\left(\var(\P{\mathrm{A}1}^{\mathrm{ent}}) +
        \var(\P{\mathrm{A}2}^{\mathrm{ent}})\right) \\
      < 1+\kappa^2 &= \var(A_1) + \var(B_1).
  \end{split}
\end{equation}
Simply stated, we must predict the outcomes $A_2$ and $B_2$ with a
precision better than the statistical spreading of the outcomes $A_1$
and $B_1$ with the additional constraint that $A_1$ and $B_1$ are
outcomes of quantum noise limited measurements.
\subsubsection{Theoretical Entanglement Modeling}
\label{sec:theo_ent_model}
We described above the experimental procedure for generating and
verifying the entangled states. Here we present a simple way to derive
what we expect for the mean values (i.e.~the $\alpha$-parameter) and
for the variances $\var(\P{\mathrm{A}i}^{\mathrm{ent}})$.

We calculate directly the expected conditional variance of $A_2$ based
on $A_1$:
\begin{equation}
  \begin{split}
    &\mean{\left(\X{\mathrm{L}1}^{\mathrm{out,2nd}} -
                 \alpha\X{\mathrm{L}1}^{\mathrm{out,1st}}\right)^2} \\
   = &\mean{\left(\X{\mathrm{L}1}^{\mathrm{in,2nd}} -
                  \alpha\X{\mathrm{L}1}^{\mathrm{in,1st}}
      + \kappa\left[\P{\mathrm{A}1}^{\mathrm{in}} -
        \alpha\P{\mathrm{A}1}^{\mathrm{ent}}\right]\right)^2} \\
   = &\frac{1}{2}(1+\alpha^2 + \kappa^2(1-\alpha)^2).
  \end{split}
\end{equation}
In the second step we assume that a perfect QND measurement without
any decoherence is performed, i.e.~$\P{\mathrm{A}1}^{\mathrm{ent}} =
\P{\mathrm{A}1}^{\mathrm{in}}$. By taking the derivative with respect
to $\alpha$ we obtain the theoretical minimum
\begin{equation}
\begin{split}
\label{eq:var_theory}
  \var(A_2|A_1) + \var(B_2|B_1) &= 1 + \frac{\kappa^2}{1+\kappa^2} \\
  \Rightarrow  \var(\P{\mathrm{A}1}^{\mathrm{ent}})+\var(\P{\mathrm{A}2}^{\mathrm{ent}})
  &= \frac{1}{1+\kappa^2}
\end{split}
\end{equation}
obtained with the $\alpha$-parameter
\begin{equation}
\label{eq:alpha_theory}
  \alpha = \frac{\kappa^2}{1+\kappa^2}.
\end{equation}
It is interesting that, in principle, any value of $\kappa$ will lead to
creation of entanglement. The reason for this is our prior knowledge
to the entangling pulse. Here the atoms are in a coherent state
which is as well defined in terms of variances as possible for
separable states. We only need an ``infinitesimal'' extra knowledge
about the spin state to go into the entangled regime.

It is also interesting to see what happens to the conjugate variables
$\X{\mathrm{A}i}$ in the entangling process. This is governed by
Eq.~(\ref{interact_atom}). We do not perform measurements of the light
operator $\P{\mathrm{L}i}^{\mathrm{in}}$ so all we know is that both
$\X{\mathrm{A}i}^{\mathrm{in}}$ and $\P{\mathrm{L}i}^{\mathrm{in}}$
are in the vacuum state. Hence $\var(\X{\mathrm{A}i}^{\mathrm{ent}}) =
(1+\kappa^2)/2$ and we preserve the minimum uncertainty relation
$\var(\X{\mathrm{A}i}^{\mathrm{ent}})\var(\P{\mathrm{A}i}^{\mathrm{ent}})
= 1/4$.

\subsubsection{Entanglement Model With Decoherence}
Practically our spin states decohere between the light pulses and also
in the presence of the light. We model this decoherence naively by
attributing the entire effect to the time interval between the two
pulses, i.e.~we assume there is no decoherence in presence of the
light but a larger decoherence between the pulses. We may then perform
an analysis in complete analogy with the above with the only
difference that $\P{\mathrm{A}1}^{\mathrm{ent}} =
\beta\P{\mathrm{A}1}^{\mathrm{in}}+\sqrt{1-\beta^2}\vac{p}$ where
$\vac{p}$ is a vacuum operator admixed such that $\beta = 0$
corresponds to a complete decay to the vacuum state and $\beta = 1$
corresponds to no decoherence. Completing the analysis we find the
theoretical conditional variances
\begin{equation}
\begin{split}
\label{eq:var_theory_decoh}
  \var(A_2|A_1) + \var(B_2|B_1) &= 1 +
    \kappa^2\frac{1+(1-\beta^2)\kappa^2}{1+\kappa^2} \\
  \Rightarrow  \var(\P{\mathrm{A}1}^{\mathrm{ent}})+\var(\P{\mathrm{A}2}^{\mathrm{ent}})
  &= \frac{1+(1-\beta^2)\kappa^2}{1+\kappa^2}
\end{split}
\end{equation}
obtained with $\alpha$-parameter
\begin{equation}
\label{eq:alpha_theory_decoh}
  \alpha = \frac{\beta\kappa^2}{1+\kappa^2}.
\end{equation}
In the limit $\beta \rightarrow 1$ these results agree
with~(\ref{eq:var_theory}) and~(\ref{eq:alpha_theory}). For $\beta
\rightarrow 0$ we have $\alpha \rightarrow 0$ (outcomes $A_1$ and
$B_1$ are useless) and the variance approaches that of the vacuum
state which is a separable state.
\subsubsection{Gaussian State Modeling}
Following the extensive study of the evolution of Gaussian states
during arbitrary interactions and measurements~\citep{giedke:02,
  eisert:03} the development of spin squeezing in a single atomic
sample and entanglement in two samples were treated in detail
theoretically in~\citep{madsen:spinsq, sherson:gaussent}. Arbitrary
Gaussian states of $n$ canonical modes are fully characterized by a
$2n\times 1$ vector, $\mathbf{v}$, describing the mean values and a
$2n\times 2n$ matrix, $\gamma$, describing the correlations within the
atomic and light systems and the cross-correlations between these. In
this formalism an interaction between light and atoms is governed by:
$\mathbf{v}\to S\mathbf{v}$ and $\gamma\to S\gamma S^T$. For the
Faraday interaction, the coefficients of the matrix $S$ are easily
determined from Eqs.~(\ref{eq:main_equations}). Decoherence is easily
included but most importantly, there is an explicit expression for the
state of the remaining modes after an arbitrary homodyne measurement
on a number of the modes is performed. If the incoming pulse of
light is split into a large number of segments and the interaction
with atoms and subsequent measurement of each segment is treated
sequentially differential equations for $\mathbf{v}(t)$ and
$\gamma(t)$ can be obtained in the limit of infinitesimal segment
durations. The former is a stochastic differential equation determined
by the outcome of the measurement, whereas the latter is
deterministic, although non-linear because of the measurement
dynamics. In~\citep{madsen:spinsq, sherson:gaussent} such differential
equations are solved giving the time resolved dynamics in the presence
of e.g.  atomic decoherence and light losses. In this way, e.g.
analytic expressions for the optimal degree of spin squeezing and the
degree of entanglement at arbitrary rotation frequencies (the
verification and generalization of Eq.~(\ref{eq:var_theory})) can
be obtained easily.
\subsection{Quantum Memory}
\label{sec:quantum-memory-protocols}
For complete quantum memory we require 1) that the light state to be
stored is supplied by a third party in an unknown state, 2) that this
state is mapped onto an atomic state with a fidelity higher than the
best classical fidelity, and finally 3) that the stored state can be
retrieved from memory. As is described in
Sec.~\ref{sec:quant-map-results} the first two criteria have been met
experimentally in \citep{ourNature04} whereas the last one still
remains an unsolved experimental challenge. A recently developed
experimentally feasible protocol for retrieval now exists and is 
discussed in Sec. \ref{sec:retrieval}.
\subsubsection{Direct Mapping Protocol}
In Eqs.~(\ref{interact_light}-\ref{interact_atom}) one of the light
variables is mapped onto one of the atomic variables.  This represents
a natural starting point for a quantum memory protocol in which the
entire light mode described by the two non-commuting variables
$\xh^{\mathrm{in}}$ and $\ph^{\mathrm{in}}$ is faithfully stored. In
the so-called "direct mapping protocol" of \citep{ourNature04} the
mapping is completed by measuring the remaining light quadrature
$x^{\mathrm{out}}=x^{\mathrm{in}}+\kappa P^{\mathrm{in}}$ and feeding
the result back into the atomic $\X{\mathrm{A}}$ with a gain of $g$:
\begin{subequations}
  \label{eq:mapping}
  \begin{align}
    \X{\mathrm{A}}^{\mathrm{out}} &= \X{\mathrm{A}}^{\mathrm{in}} +
       \kappa \P{\mathrm{L}}^{\mathrm{in}},\\
    \P{\mathrm{A}}^{\mathrm{out}}~'&= \P{\mathrm{A}}^{\mathrm{out}} -
    g\X{\mathrm{L}}^{\mathrm{out}} = \P{\mathrm{A}}^{\mathrm{in}}(1-\kappa g) -
      g \X{\mathrm{L}}^{\mathrm{in}}.
  \end{align}
\end{subequations}
If $\kappa=g=1$ and the initial atomic state is assumed to be a
coherent state with zero mean value the mean values of both light
variables will be stored faithfully in the atoms. Although the initial
atomic state has zero mean it is a quantum mechanically fluctuating
state and any uncanceled atomic part will increase the noise of the
final state and thus degrade the mapping performance. Although this
protocol works for any state, in the following we discuss storage of
coherent states of light, i.e.  vacuum states which are displaced by
an unknown amount in phase space. For the storage of an arbitrary
coherent light state the remaining $\X{\mathrm{A}}^{\mathrm{in}}$
contribution limits the storage fidelity to 82\%.  This can be
remedied by initially squeezing the atomic state, in which case 100 \%
fidelity can be reached in the limit of infinite squeezing.

\subsubsection{Mapping With Decoherence}
Just as in the case of entanglement generation the spin states 
decohere. Again we can model this by a beam splitter type admixture of
vacuum components right after the passage of the first light pulse. We
can furthermore model light damping (e.g. reflection losses) in a
similar way to obtain:
\begin{subequations}
\label{eq:mapdecoher}
\begin{align}
  \X{\mathrm{A}}^{\mathrm{out}} &\to \beta(\X{\mathrm{A}}^{\mathrm{in}} +
     \kappa \P{\mathrm{L}}^{\mathrm{in}}) + \sqrt{1-\beta^2} V_{X\mathrm{A}}, \\
  \P{\mathrm{A}}^{\mathrm{out}} &\to (\beta-g\kappa\sqrt{\zeta})
     \P{\mathrm{A}}^{\mathrm{in}}-g\sqrt{\zeta}\X{\mathrm{L}}^{\mathrm{in}} +
     \sqrt{1-\beta^2}V_{P\mathrm{A}}-g\sqrt{1-\zeta}V_{X\mathrm{L}}~.
\end{align}
\end{subequations}
We see that $\P{\mathrm{L}}^{\mathrm{in}}$ and
$\X{\mathrm{L}}^{\mathrm{in}}$ are mapped with gains
$g_{\mathrm{BA}}'=\beta\kappa$ and $g_{\mathrm{F}}'=g\sqrt{\zeta}$
respectively. The variances can be calculated easily to be:
\begin{subequations}
\label{eq:vardecoher}
  \begin{align}
    \mathrm{Var}(X^{\mathrm{out}}) &= 1+g_{\mathrm{BA}}'^2,\\
    \mathrm{Var}(P^{\mathrm{out}}) &= 1+\frac{g_{\mathrm{F}}'^2}{\zeta} +
      \frac{g_{\mathrm{F}}'^2g_{\mathrm{BA}}'^2}{\beta^2} - 2g_{\mathrm{F}}'g_{\mathrm{BA}}'.
  \end{align}
\end{subequations}
\subsubsection{Quantum Memory Retrieval}
\label{sec:retrieval}
As mentioned in \citep{ourNature04} the stored state can in principle
be retrieved by inverting the roles of light and atoms in the direct
mapping protocol. This would involve first an interaction between a
read-out light beam with the atomic sample acting as a storage medium.
According to Eq.~(\ref{interact_light}) this would map
$\P{\mathrm{A}}$ onto the light. Next $\X{\mathrm{A}}$ has to be
measured and feedback applied to the read-out beam according to the
result of the measurement. However, since the atomic measurement
requires a certain time during which the read-out pulse propagates at
the speed of light, the feedback is only practical for pulse durations
shorter than a microsecond. In the experiments of~\citet{ourNature04}
pulses of millisecond duration ($\sim 300\kilo\meter$) are required in
order to obtain a sufficiently high interaction strength, and the
inverse direct mapping protocol is thus infeasible for this
experimental realization.

Several years ago a retrieval scheme, which did not involve
measurements, but instead used two orthogonal passages of the read-out
pulse was proposed \citep{contvarbook}. In each passage one of the
atomic quadratures is mapped onto the light pulse, and in this way
retrieval fidelities of up to 82\% can be achieved with coherent
readout light and up to 100\% with perfectly squeezed readout light.
Unfortunately in order to preserve the QND nature of each of the two
interactions the light has to pass entirely through the atomic medium
before proceeding to the second passage.  This again renders the
protocol inapplicable to all setups requiring "long" pulses.

Recently this problem was eliminated by solving the complex dynamics
arising when a light beam passes through the atomic medium along two
orthogonal directions simultaneously~\citep{ourMultiPass}. This
protocol works with arbitrary pulse durations and the fidelity of this
two-pass protocol has been calculated both for a coherent input state
and for a light qubit. It was also shown that if the light is
reflected back after the second passage, thus completing four
passages, and a time dependent interaction strength is applied perfect
retrieval can be achieved without requiring squeezed initial states.
In a related proposal~\citep{fiurasek:retrieval} retrieval is achieved
by sending two different beams through the atomic samples
simultaneously, each in a direction orthogonal to the other. The
fidelity of retrieval for coherent states including realistic light
losses is calculated and shown to exceed the classical bound.
\subsubsection{Alternative Quantum Memory Proposals}
\label{sec:altern-quant-memory}
In~\citep{contvarbook} various protocols which can be implemented
using the QND-Faraday interaction are described in detail, including
atom-atom teleportation. Instead of reviewing these, we would like to
mention some recent theoretical proposals for improving the quantum
memory performance. All of these break with the simple interaction
scheme either by introducing a non-QND interaction (as was also done
in~\citep{ourMultiPass, fiurasek:retrieval}) or by exploiting the fact
that the atoms involved are not simple spin-1/2 atoms and thus contain
more than two magnetic sublevels.

In~\citep{opatrny:enhcapqmem} it is proposed to exploit additional
atomic coherences to enhance the capacity of an atomic quantum memory.
Remember that during the usual QND Faraday interaction $\Delta m=1$
atomic coherences are coupled to the sideband of the light at the
Larmor frequency. If in addition circularly polarized classical fields
are present during interaction coupling between $\Delta m=2$
coherences and a light sideband at twice the Larmor frequency are also
created. In this way an additional quantum channel is added to the
quantum memory, thus enhancing its capacity. By appropriately tuning
the detuning of the additional coupling field two-mode squeezer and
beam splitter Hamiltonians can also be realized. With the latter, a
quantum memory is realized in a single passage without a requirement
of measurements or prior squeezing of atoms or light.

In~\citep{opatrny:singlecellqmem} the author proposes to create a
single cell atomic memory by optically pumping the sample into an
incoherent mixture of the two extreme magnetic sublevels ($m=-4$ and
$m=4$ for the $F=4$ hyperfine level of cesium). This is done with
suitable linearly polarized light. When sending a light pulse through
the atomic sample it will interact with both extreme coherences
simultaneously, creating the usual QND-Faraday interaction. Apart from
the prospect of making the memory more compact, this proposal would
also substantially decrease the reflection losses associated with the
many air-glass transitions in a two-cell setup. A main problem in this
approach is that in the presence of the probe light the two coherences
will experience different AC-Stark shifts and thus different phase
evolutions. For relevant pulse durations this effect is in fact
significant. The author proposes to solve this by either decreasing
the bias magnetic field or by introducing an additional AC-Stark shift
by adding an light field with appropriate polarization and detuning.

An atomic memory could of course also be implemented using light-atom
teleportation. This could be achieved by first sending an auxiliary
pulse through the two oppositely oriented atomic samples. This would
entangle the light and atomic systems via the QND-Faraday interaction.
After this, the auxiliary light beam should be mixed on a 50/50 beam
splitter with the quantum light signal to be teleported.  Measurements
of $S_y$ and $S_z$ in the two output ports respectively and subsequent
feedback to the atomic system would complete the teleportation.
Unfortunately the achievable fidelity is limited to 67\% with the
usual QND-interaction. It was therefore proposed
in~\citep{hammerer:teleportation} to replace the two atomic samples in
the protocol just sketched by a single sample, still with a constant
bias magnetic field. This changes the interaction dramatically. Both
atomic quadratures are transferred to the light, albeit at the cost of
additional new modes of light. The resulting multi-mode light-atom
entanglement enables teleportation with up to 77\% fidelity.

%
%
\section{Experimental Methods}
\label{sec:exp-setup}
\begin{figure}[t]
\centerline{\includegraphics[width=\linewidth]{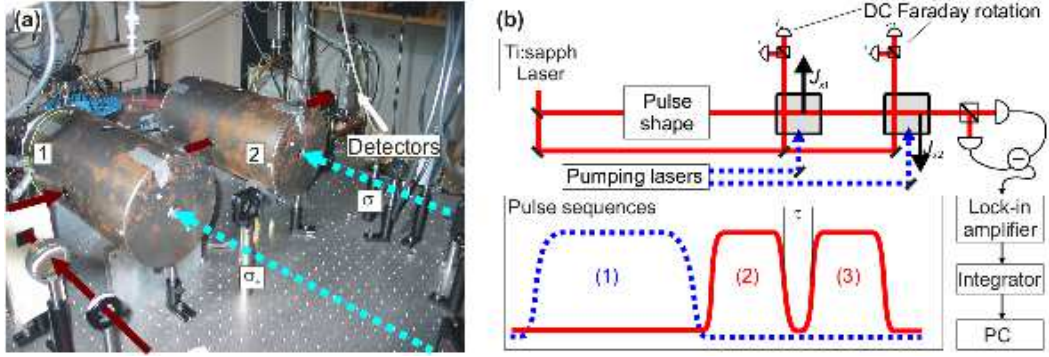}}
\caption{\small {\bf(a)} 
  A photographic view of a typical experimental setup. Atomic vapor
  cells are placed inside the cylindrical magnetic shields. The
  pumping beams are indicated with dashed arrows and the path of the
  quantum probe field is marked with the solid arrows.  {\bf (b)} A
  schematic view of the setup. The probe pulses reach a detection
  system measuring $\S{y}(t)$. The photo current is sent to a lock-in
  amplifier which singles out the $\sin(\Omega t)$ and $\cos(\Omega
  t)$ parts as we discuss in Sec.~\ref{sec:detect_pol_states}. These
  are integrated and stored in a PC. Typically, the pulse sequence
  consists of (1) a pumping pulse, (2) and (3) two laser pulses for
  quantum manipulation and detection of the atomic states.A very small
  portion of the probe field is sent through each sample in the
  $x$-direction to measure the magnitude $J_x$ of the macroscopic
  spins by Faraday rotation measurements.}
 \label{fig:exp_setup_typical}
\end{figure}
In this section we describe the typical setup for our experiments.  It
is centered around two glass cells filled with Cs vapor at room
temperature placed in two separate magnetically insulating shields
with a bias magnetic field inside. Additional coils are used to apply
an rf magnetic field with the frequency equal to the Larmor frequency
of the bias field. Atoms in the cells are optically pumped. The setup,
therefore, is similar to that of a classical magneto-optical resonance
experiment.

The typical experimental setup is shown in
Fig.~\ref{fig:exp_setup_typical}. A Verdi V8 pumped Ti:sapphire laser
delivers what we call the probe field. This is used for quantum
manipulation of the atomic samples and also for detecting the
macroscopic spin $J_x$. Diode lasers provide optical pumping
fields for creating highly polarized spin states. The cw lasers are
modulated by AOMs or EOMs to achieve the desired pulse shapes. Atoms
are contained in vapor cells and placed in stable magnetic fields.
The magnetic field homogeneity must be of order $10^{-3}$ across
the vapor cell volume.

\subsection{Paraffin Coated Vapor Cells}
\label{sec:vapour_cells}
In our experiments the atomic samples are contained in a paraffin
coated vapor cell. The coating prevents depolarization of the spin
state when atoms hit the walls. We have measured spin coherence times
exceeding 40ms and spin polarizations exceeding 99\% by the methods
described in Sec.~\ref{sec:char-spin-state}. The atomic density and
thereby the macroscopic spin $J_x$ can be controlled by heating
or cooling the vapor cell. In order to achieve a stable vapor density,
temperature gradients across the cell should be avoided.  Temperature
control by air flow is a convenient solution.  Metal heating/cooling
elements causes severe problems since the atoms are disturbed
by random magnetic fields created by thermal currents even if aluminum
is used, see~\citep{julsgaard_phd}.  Further information on paraffin
coated cells can be found in~\citep{bouchiat:66, alexandrov:96,
  alexandrov:02}.
\subsection{Detection of Polarization States}
\label{sec:detect_pol_states}
\begin{figure}[t]
\centering{\includegraphics[width=10cm]{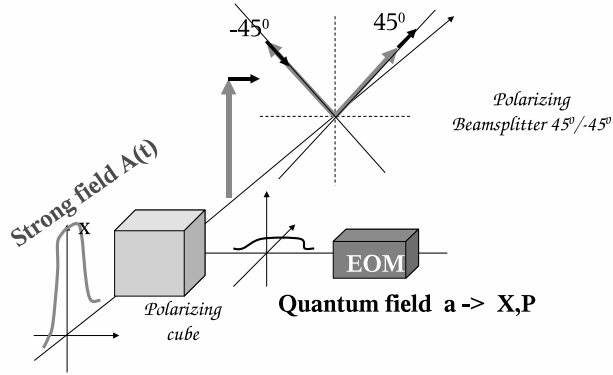}}
\caption{\small 
  The setup for measuring $\S{y}$. A strong classical pulse is mixed
  on a polarizing beam splitter with the quantum field which can be
  in a vacuum state, as in the entanglement experiment.
  Measurement is performed in the $45^{\circ}$- and
  $-45^{\circ}$-basis by two detectors. Half of the difference of the
  two photo currents is $\S{y}$. An optional $\lambda/4$-retardation
  plate turns this measurement into the measurement of $\S{z}$
  component.}
\label{fig:meas_sy}
\end{figure}
The Stokes parameters are measured with low noise photo detectors. We
use high quantum efficiency photo diodes and home made amplifiers
characterized by negligible electronic noise compared to the shot
noise of light at optical power higher than $1 mW$.  In
Fig.~\ref{fig:meas_sy} we depict how the $\S{y}$-component of light is
measured.

The differential photo current $i(t)$ from the two detectors
corresponds to a realization of the measurement of $\S{y}$. By
passing $i(t)$ through a lock-in amplifier we can detect the sine and
cosine components at the Larmor precession frequency $\Omega$.
Practically, the current $i(t)$ is multiplied by $\cos(\Omega t)$ and
$\sin(\Omega t)$ and integrated over time for this purpose. According
to~(\ref{eq:def_XL1}-d), with appropriate scaling, this exactly
corresponds to measuring the $\X{\mathrm{L}1}$ and $\X{\mathrm{L}2}$
components of light. For $\S{z}$-detection we would measure
$\P{\mathrm{L}1}$ and $\P{\mathrm{L}2}$.

\subsection{ Magnetic Fields}
\label{sec:bias_mag_field} As was discussed in
Sec.~\ref{sec:addbfield}, external magnetic fields are added to control
the dynamics of the atomic samples. The interaction of atoms with a
magnetic field is governed by the Hamiltonian
\begin{equation}
  \label{eq:Hamilton_ex_field}
  \H_{\mathrm{mag}} = g_F\mu_{\mathrm{B}}\vec{J}\cdot\vec{B} + O(B^2)
\end{equation}
We stress here that $\vec{J}$ is the total angular momentum of the
atom including the nuclear spin. For the $F=4$ ground state of cesium
$g_F \approx 1/4$. The second term $O(B^2)$ reminds us that the above
linear equation is only approximately true. When the magnetic energy
becomes comparable to the hyperfine splitting of the ground state the
response is non-linear. We comment on this below.

Adding a constant bias magnetic field $B_x$ in the $x$-direction leads
to the equations of motion~(\ref{eq:rotating_frame_dynamics}) with
Larmor precession at the frequency $\Omega_{\mathrm{L}} = g_F
\mu_{\mathrm{B}}B_x/\hbar$.  If we furthermore add an RF magnetic
field at frequency $\Omega$ along the $y$-direction such that
\begin{equation}
  \vec{B}_{\mathrm{ext}} = B_x\vec{e}_x +
     \left(B_c \cos(\Omega t+\phi) + B_s \sin(\Omega t+\phi)\right)\vec{e}_y
\end{equation}
with constants $B_c$ and $B_s$ we may derive for the rotating frame
coordinates $\J{y}'$ and $\J{z}'$
of~(\ref{eq:rotating_frame_dynamics}) that
\begin{equation}
\label{eq:RF-field-influence-general}
    \frac{\partial\J{y}'(t)}{\partial t} =
       -\omega_s \sin(\Omega_{\mathrm{L}} t)\sin(\Omega t+\phi)J_x, \quad
    \frac{\partial\J{z}'(t)}{\partial t} =
       -\omega_c \cos(\Omega_{\mathrm{L}} t)\cos(\Omega t+\phi)J_x,
\end{equation}
with $\omega_{c,s} = g_F\mu_{\mathrm{B}}B_{c,s}/\hbar$. Choosing the
phase and the frequency of the RF-drive such that $\phi = 0$ and
$\Omega = \Omega_{\mathrm{L}}$ we obtain:
\begin{equation}
\label{eq:RF-field-influence}
    J_x(t) = J_x(0), \quad
    \frac{\partial\J{y}'(t)}{\partial t} = -\frac{\omega_s J_x}{2}, \quad
    \frac{\partial\J{z}'(t)}{\partial t} = -\frac{\omega_c J_x}{2}.
\end{equation}
These equations are valid for interaction times $T$ such that
$\omega_c T, \omega_s T \ll 1 \ll \Omega T$. We see that with pulses
of RF-magnetic fields we are able to change the spin components
$\J{y}'$ and $\J{z}'$ by an amount controlled by the sine and cosine
components $B_s$ and $B_c$. This has several experimental
applications, which are discussed below.

\subsubsection{Characterizing the Spin State with the Magneto-optical
  Resonance Method}
\label{sec:char-spin-state}

Equations~(\ref{eq:Syout_rotating_two}) and
(\ref{eq:RF-field-influence-general}) describe the Magneto-Optical
Resonance method (MORS) which we use extensively for the spin state
characterization. Application of MORS to our experiments are described
in detail in ~\citet{julsgaard:mors}.  Within this method the RF
frequency is scanned around $\Omega_{\mathrm{L}}$ and the oscillating transverse
spin components are probed via oscillating polarization rotation of
the optical probe.  In order to quantitatively explain the MORS signal
as the RF-frequency is scanned across $\Omega_{\mathrm{L}}$, we need to turn
back to the second order term mentioned in
Eq.~(\ref{eq:Hamilton_ex_field}).  The transverse spin components
$\j{y}$ and $\j{z}$ can be expressed in terms of coherences
$\dens{m,m\pm 1}$ in the following way:
\begin{equation}
\label{eq:spin_dens}
  \begin{split}
  \j{y} &= \frac{1}{2}\sum_m\sqrt{F(F+1)-m(m+1)}(\dens{m+1,m}+\dens{m,m+1}),\\
  \j{z} &=\frac{1}{2i}\sum_m\sqrt{F(F+1)-m(m+1)}(\dens{m+1,m}-\dens{m,m+1}).
   \end{split}
\end{equation}
In the absence of the second order term
in~(\ref{eq:Hamilton_ex_field}) the energy separation $\hbar\Omega_{\mathrm{L}}$
between states $\ket{m}$ and $\ket{m+1}$ is the same for all $m$ and
all terms $\dens{m+1,m}$ have the same resonant frequency.  The second
order term in~(\ref{eq:Hamilton_ex_field}), however, makes the
frequency of the coherences $\dens{m+1,m}$ slightly different. It can
be shown, that the frequency difference $\omega_{\mathrm{QZ}}$ between
$\dens{m,m+1}$ and $\dens{m-1,m}$ is $\omega_{\mathrm{QZ}} =
2\Omega_{\mathrm{L}}/\omega_{\mathrm{hfs}}$ where $\omega_{\mathrm{hfs}} =
2\pi\cdot 9.1926\mathrm{GHz}$ is the hyperfine splitting of the Cesium
ground state. We typically have $\Omega_{\mathrm{L}} = 2\pi\cdot 322\mathrm{kHz}$
and the effect is small but detectable.

\begin{figure}[t]
  \centering
\includegraphics[width=0.5\textwidth,angle=0]{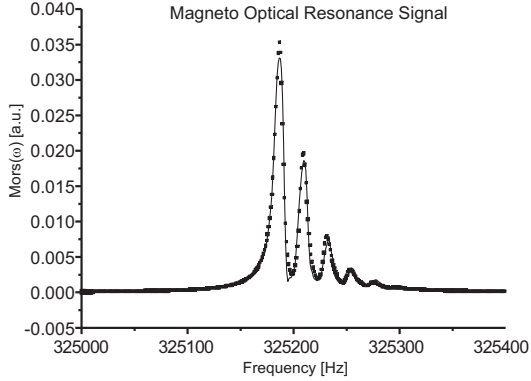}
\caption{\footnotesize 
  MORS signal, in arbitrary units (a.u.) of a poorly oriented
  sample.  Each peak corresponds to a $m_F \leftrightarrow m_F+1$
  Zeeman resonance.  The height of each resonance is proportional to
  the population difference of the two relevant sublevels. }
\label{fig:mors}
\end{figure}

In the special case that the amplitude and frequency of the driving
RF-field vary slowly compared to the spin coherence time, the
off-diagonal coherences follow the diagonal populations adiabatically
and we may write e.g.~$\J{y}$ as:
\begin{equation}
\label{eq:mors}
  \J{y} = \mathrm{Re}\left[\mathrm{const}\sum_{m=-F}^{F-1}\frac{[F(F+1)-m(m+1)]
          \cdot e^{i\Omega t}}{i(\Omega_{m+1,m}-\Omega)-\Gamma_{m+1,m}/2}
          [\hat{\sigma}_{m+1,m+1}-\hat{\sigma}_{m,m}]\right]
\end{equation}
where $\Gamma_{m+1,m}$ are the FWHM linewidths giving an exponential
$e^{-\Gamma t/2}$ decay of each coherence. The Larmor frequency
$\Omega_{\mathrm{L}}$ has been replaced by the individual coherence evolution
frequencies $\Omega_{m+1,m}$. For $\J{z}$ we have to take the
imaginary part. Two adjacent magnetic sublevels act as a two level
system with the usual Lorentzian response to a driving RF field.
Scanning the RF frequency we get eight Lorentzian peaks, the
magnitudes of which will depend on the populations of the magnetic
sublevels. The MORS signal is proportional to the square of the term
in the square brackets in Eq.~(\ref{eq:mors}). An example of such a
signal can be seen in Fig.~\ref{fig:mors}. The application of this
method is twofold. First, we gain information on the
distribution of population among the different ground state magnetic
sublevels. From this we infer that we are able to optically pump the
atoms to such an extent that only the outermost coherence
($m_F=4\leftrightarrow m_F=3$) becomes significant. Second,
measuring the widths of the resonances under different experimental
conditions allows us to quantify the effect of different decoherence
mechanisms as discussed in Sec.~\ref{sec:decoherence}.

A second implication of the second order term is that the equations of
motion $\partial\j{y}/\partial t = -\Omega\j{z}$ and
$\partial\j{z}/\partial t = \Omega\j{y}$ describing Larmor precession
are too simple and must in principle be generalized.  However, this is
not necessary for our case with highly polarized samples. We have
almost all the population in the $m_F = 4$ and $m_F = -4$ states for
the two samples. Then there is effectively only one non-vanishing
frequency component in the sums~(\ref{eq:spin_dens}).  For the $j_x =
4$ sample the single term is $\dens{3,4}$ and for the $j_x = -4$
sample the only term is $\dens{-3,-4}$.

\subsubsection{Manipulating the Spin State}
\label{sec:manipulating_spin_state}
Of course, as suggested by Eq.~(\ref{eq:RF-field-influence}),
an appropriate choice of phase, strength and envelope function
of the RF-field allows us to create an arbitrary mean value of either
of the spins or any combination of these. This has two applications.
First, it enables us to create a large classical mean value and
observe the evolution of this state, thus constituting another
mechanism for studying decoherence and calibrating the system as is
discussed briefly at the end of Sec.~\ref{sec:polarization_and_Stark}.
Second, we can actively feed back the result of a quantum probing of
the spin state, thus creating a particular desired state. This is the
keystone element for creation of deterministic entanglement, discussed
in Sec.  \ref{subsec:uncondEnt}, and for the quantum mapping
experiment discussed in Sec. \ref{sec:quant-map-results}.

\section{Experimental Results}

%
%
\subsection{Projection Noise Level}
\label{sec:proj-noise-level}
Since the Heisenberg uncertainty relation sets the starting point of
all our calculations, one of the most important tasks in our
experiments is the achievement of quantum noise limited
performance. Practically, it is also one of the most difficult tasks.
When we detect polarization states of light we observe noise in the
signals.  After the light has passed the atomic samples, there is
a contribution to this noise from the light itself and from the atomic
spins. The noise contribution from atoms in the minimum uncertainty
state (the coherent spin state) is called the projection noise.

We discussed the ratio of the projection noise to the quantum noise of
light (shot noise) already in Sec.~\ref{sec:ent_gen_ver}. We
found that theoretically this ratio should be
\begin{equation}
  \label{eq:kappa_sqr_naive}
  \kappa^2 = a^2 J_x S_x T.
\end{equation}
The ratio $\kappa^2$ came from the more general interaction
equations~(\ref{interact_light}-d). In the present section we discuss
how to calibrate this projection noise level experimentally and how to
predict the noise level from independent measurements.

\subsubsection{Measuring the Macroscopic Spin}
\label{sec:meas_macro_spin}
The ratio of projection noise to shot noise is proportional to the
macroscopic spin $J_x$. This linearity is the finger print of
quantum noise and is essential to establish experimentally. To this end
we need a good measure of $J_x$. 

We measure $J_x$ by detecting polarization rotation of a linearly
polarized laser field propagating through the atomic samples along the
$x$-direction. To see what happens in this setting we consider
Eqs.~(\ref{eq:propagation_spin}) and (\ref{eq:propagation_light}). These
equations assume propagation along the $z$-direction so we assume the
spin to be polarized along the $z$-direction in the following. For
linearly polarized light we have $\mean{\S{z}} = 0$ and the effect on
the transverse spin components $\j{x}$, $\j{y}$ is negligible. It can
also be shown easily that the $a_0$ and $a_2$ terms of the
Hamiltonian~(\ref{sec_inter:final_eff_Hamil_real}) play no role in
this calculation. We are left with the evolution of $\S{x}$ and
$\S{y}$ according to~(\ref{eq:propagation_light}) and after
integration over the sample we find
\begin{equation}
  \begin{split}
    S_x^{\mathrm{out}} &= S_x^{\mathrm{in}}\cos(2\theta_{\mathrm{F}}) 
                          - S_y^{\mathrm{in}}\sin(2\theta_{\mathrm{F}}), \\
    S_y^{\mathrm{out}} &= S_x^{\mathrm{in}}\sin(2\theta_{\mathrm{F}}) 
                          + S_y^{\mathrm{in}}\cos(2\theta_{\mathrm{F}}),
  \end{split}
\end{equation}
where ``in'' refers to the polarization state before the sample at
$z=0$ and ``out'' refers to the state after the sample at $z=L$. The
angle $\theta_{\mathrm{F}}$ is given by (in radians)
\begin{equation}
  \label{eq:faraday_rotation}
  \theta_{\mathrm{F}} = -\frac{a_1\gamma\lambda^2\rho L}{32\pi\Delta}\cdot
     \mean{\j{z}}. 
\end{equation}
If a linearly polarized beam of light is rotated by the angle
$\theta$, the Stokes vector is rotated by $2\theta$. Thus, in the
above, $\theta_{\mathrm{F}}$ is the polarization rotation caused by
the spin orientation along the direction of light propagation. We note
that the angle $\theta_{\mathrm{F}}$ depends on the density $\rho$ of
atoms times the length $L$ that the light traverses. We wish to
re-express this in terms of the macroscopic spin size $J_z =
N_{\mathrm{at}}\mean{\j{z}}$ of the entire sample (remember we have
the spins polarized along $z$ in this discussion). To this end we
observe that $N_{\mathrm{at}} = \rho V \equiv \rho A_{\mathrm{cell}}
L$ where $V$ is the vapor cell volume and $A_{\mathrm{cell}}$ is the
area of the vapor cell transverse to the beam direction.  This will
conveniently be an effective area for cells that are not exactly box
like. Turning back to the usual convention of spin polarization along
the $x$-axis we then rewrite~(\ref{eq:faraday_rotation}) as
\begin{equation}
\label{eq:faraday_angle}
  \theta_{\mathrm{F}} = -\frac{a_1\gamma\lambda^2 J_x}{32\pi A_{\mathrm{cell}}\Delta}.  
\end{equation}

\subsubsection{Predicting the Projection Noise Level}
\label{sec:predict_proj_noise}
Now, let us return to the predicted ratio of projection to shot
noise~(\ref{eq:kappa_sqr_naive}). This prediction relies on
Eqs.~(\ref{eq:syout}-\ref{eq:jzout}) which are derived under the
assumption that all atoms in the sample are seen by the laser beam
which has a cross sectional area $A$. In experiments the laser
beam does not intersect all the atoms. In
App.~\ref{sec:atom_motionAll} we show that the random motion of atoms
in and out of the beam modifies the expected variance of the
transverse spin components $\J{y}$ and $\J{z}$ by statistical effects
from the usual $J_x/2$ to $p^2(1+\sigma^2)J_x/2$ where $p =
A/A_{\mathrm{cell}}$ is the mean time of an atom inside the laser beam
and $\sigma^2$ is the relative variance of $p$. We furthermore present
a simple model for $p$ and $\sigma$ and show that the atomic motion
will act as an effective source of decoherence between two probe
pulses. We incorporate atomic motion into
Eq.~(\ref{eq:kappa_sqr_naive}) by replacing $A$ with
$A_{\mathrm{cell}}$ in the factor $a$ and multiplying the whole
expression by $1+\sigma^2$. We then find
\begin{equation}
  \begin{split}
  \label{eq:kappa_sqr_real}
  \kappa^2 &= a^2 J_x S_x T\cdot p^2(1+\sigma^2) = 
     \left(\frac{\gamma}{8A_{\mathrm{cell}}\Delta}\frac{\lambda^2}{2\pi}a_1\right)^2
     J_x S_x T (1+\sigma^2) \\
    &= \frac{(1+\sigma^2)\gamma\lambda^3 a_1 P\cdot T\cdot \theta_{\mathrm{F}}}
       {32\pi^2 A_{\mathrm{cell}}\Delta\hbar c}.
  \end{split}
\end{equation}
In the last step we replaced $S_x = \phi/2 = P/2\hbar\omega =
P\lambda/4\pi\hbar c$ where $P$ is the optical power. We also inserted
Eq.~(\ref{eq:faraday_angle}) to express $\kappa^2$ as a function of
$\theta_{\mathrm{F}}$. However, we must remember that the area
$A_{\mathrm{cell}}$ in~(\ref{eq:faraday_angle}) refers to the
transverse area for a beam propagating in the $x$-direction while the
$A_{\mathrm{cell}}$ from the relation $p = A/A_{\mathrm{cell}}$ refers
to the transverse area for a beam along $z$. Hence, the last step
above is valid for a vapor cell of cubic symmetry only, but it can
still be an irregularly shaped cell. In other cases the generalization
is straightforward. We have reached an expression for $\kappa^2$ in
terms of convenient parameters from an experimental point of view.
With $\gamma/2\pi  = 5.21\mathrm{MHz}$ and $\lambda = 852.3\mathrm{nm}$
together with $\hbar c$ we reach our final theoretical estimate for
the projection to shot noise ratio expressed in convenient units:
\begin{equation}
\label{eq:PN_theory_convenient}
  \kappa^2_{\mathrm{th}} = \frac{56.4\cdot P[\mathrm{mW}]
    \cdot T[\mathrm{ms}]\cdot\theta_{\mathrm{F}}[\mathrm{deg}]\cdot a_1(\Delta)
\cdot(1+\sigma^2)}{A_{\mathrm{cell}}[\mathrm{cm}^2]\cdot\Delta[\mathrm{MHz}]}, 
\end{equation}
where $a_1(\Delta)$ was defined in Eq.~(\ref{def:a0a1a2_F=4}).

\subsubsection{Experimental Investigation}
\label{sec:exp_investigation_PN}
\begin{figure}[t]
  \centering
  \begin{minipage}[t]{.45\textwidth}
  \centering
  \includegraphics[angle=0,width=\textwidth]{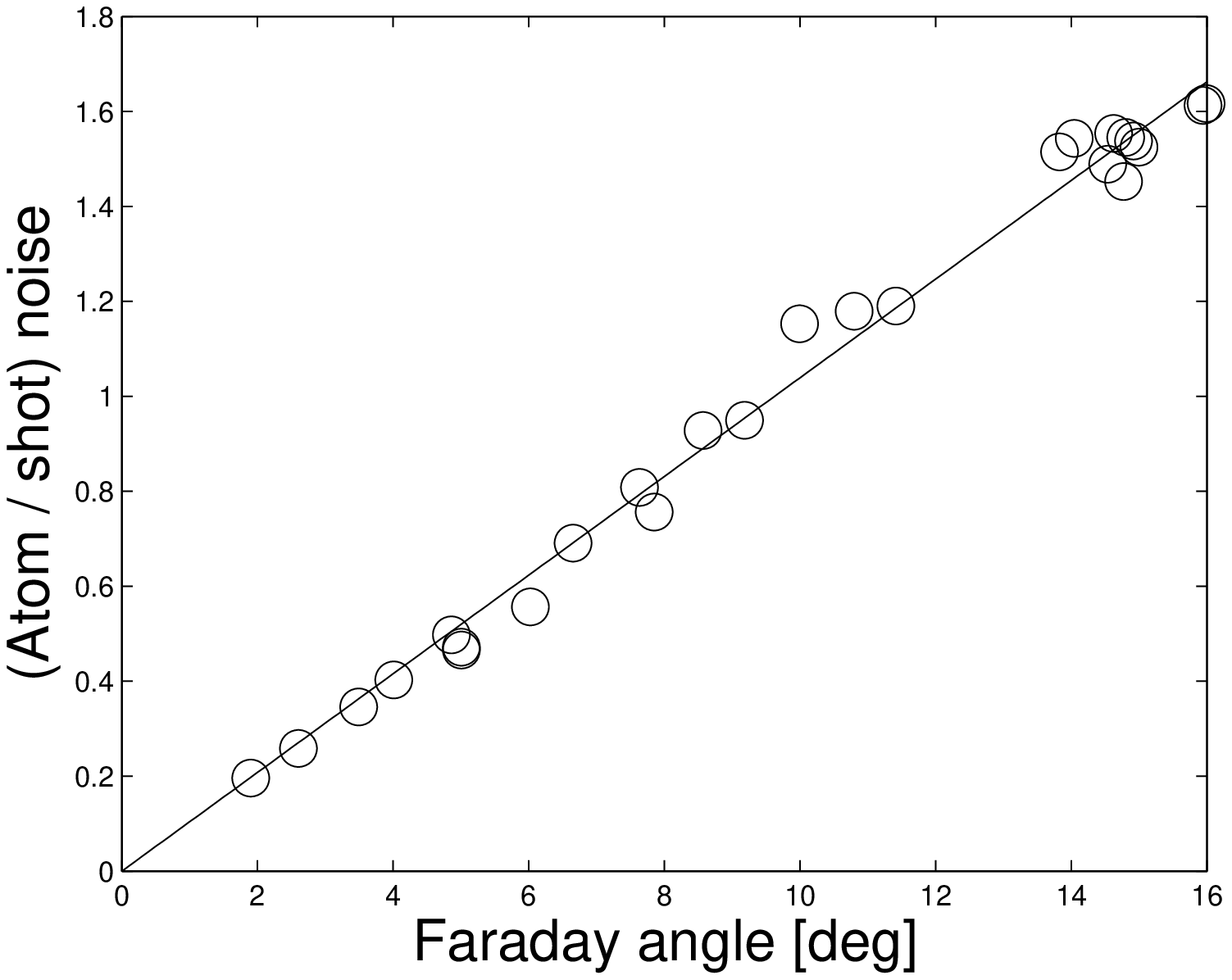}
  \caption{\footnotesize 
    Measured atomic noise relative to shot noise of light. The
    linearity is a clear signature of the projection noise limitation.
    The slope $\kappa_{\mathrm{exp}}^2 =
    0.104(2)\cdot\theta_{\mathrm{F}}$ should be compared to the
    theoretical value of $\kappa_{\mathrm{th}}^2 =
    0.140\cdot\theta_{\mathrm{F}}$ from
    Eq.~(\ref{eq:PN_theory_convenient}) with $\sigma^2 = 0$.
}
  \label{fig:proj_noise}
 \end{minipage}%
\hspace{0.5cm}%
  \begin{minipage}[t]{.45\textwidth}
    \centering
    \includegraphics[angle=0,width=\textwidth]{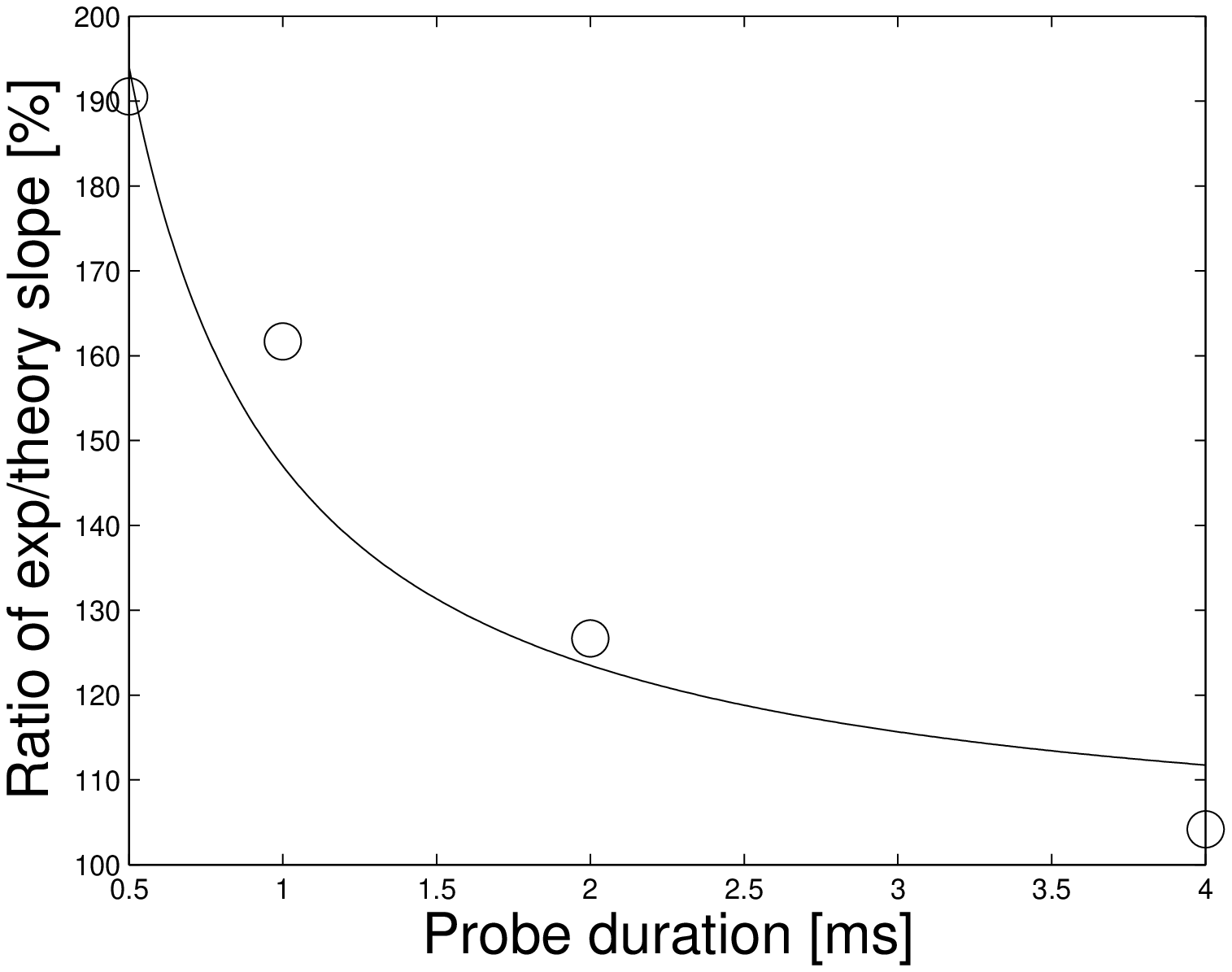}
    \caption{ \footnotesize Slope of measured $\kappa^2$ vs. $\theta$
      normalized to the experimentally predicted level (without the
      factor $1+\sigma^2$) vs. $T_{\mathrm{probe}}$. The fit gives
      $\kappa^2_{\mathrm{exp}}/\kappa^2_{\mathrm{th}}=
      1+0.47(13)/T[\mathrm{ms}]$.}
  \label{fig:sigma}
\end{minipage}
\end{figure}
Turning to experiment, in Fig.~\ref{fig:proj_noise} we see an example
where the measured noise relative to the shot noise of light is
plotted. The data are clearly linear. With $\Delta/2\pi = 700$MHz, $T
= 2.0$ms, $P = 4.5$mW, and $\sigma^2 = 0$ for the moment, we predict a
linear slope of 0.140 which is somewhat higher than the measured value
of 0.104.  Including the $\sigma^2$ from atomic motion makes the
discrepancy slightly worse but given the simplicity of the theoretical
model we consider the agreement satisfactory.

To test the scaling properties predicted in the atomic motion
calculations, we fix the power $P$, detuning $\Delta$ and macroscopic
spin size $J$ but vary the probe duration $T$. The measured noise is
plotted in Fig.~\ref{fig:sigma} relative to the
prediction~(\ref{eq:PN_theory_convenient}) with $\sigma^2=0$. We see
that as $T$ is increased we do see a lower and lower noise level which
corresponds to decreasing $\sigma^2$. The solid line in the figure
represents a fit where $\sigma^2 = (0.47\pm 0.13)/T[\mathrm{ms}]$. To
compare this to the simple model described
in~(\ref{eq:simple_model_mean_var}) we estimate our beam diameter to
be 1.6cm which gives $A \approx 2.0$cm; moreover, $L = 3.0$cm, $v_0 =
13.7$cm/ms (cesium at room temperature).  For $T = 1$ms we get the
prediction $\sigma^2 = 0.44$. This is in very good agreement with the
measured data, but this agreement must be viewed as
fortuitous. As mentioned before, numerical simulations of
atomic motion have shown that the variance
estimate~(\ref{eq:simple_model_mean_var}) is almost four times too
high. The high experimental value must be attributed to the additional
Doppler broadening effect. We also note the relatively high
uncertainty of 0.13. But all together we have a qualitative
understanding of the physics and a quantitative agreement within a few
tens of percent.
\subsubsection{Thermal Spin Noise}
\label{sec:thermal_spin_noise}
\begin{figure}[t]
  \centering
  \includegraphics[angle=0,width=8cm]{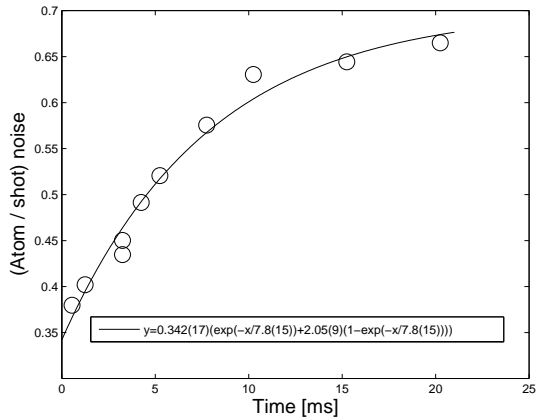}
  \caption{Coherent state noise compared to the completely unpolarized
    spin noise. The data is taken with a vapor cell in which the spin
    life time is very short. The noise level increases on a time scale
    of roughly 8ms to the thermal equilibrium level. The increase in
    noise is consistent with predictions for the coherent and unpolarized
    spin states.} 
\label{fig:unpol}
\end{figure}
Another issue concerning the projection noise level is the question of
thermal spin noise. For the establishment of the correct noise level
we must be in the CSS with high precision. For the CSS the spin is
completely polarized along the $x$-direction and $\var(j_y) =
\var(j_z) = F/2 = 2$ for the $F=4$ ground state. As a very different
example we may consider a completely unpolarized sample. We then have
by symmetry $\var(j_x) = \var(j_y) = \var(j_z) = (j_x^2+j_y^2+j_z^2)/3
= F(F+1)/3 = 20/3$. This is a factor of 10/3 higher and, even for 
fairly good polarization, the thermal noise may be significant. In our
experiments with quantum information protocols we exceed a spin
polarization of 99\% which means that the thermal noise must be very
small compared to the true projection noise. The degree of spin
polarization has been measured independently with methods similar to
those in~\cite{julsgaard:mors}. 

A nice illustration of the fact that we get lower noise for the CSS
than in the unpolarized case is given in Fig.~\ref{fig:unpol}. 
Experimentally, we perform measurements on very poor vapor cells where
the macroscopic spin life time is small. We optically pump the sample
and wait for some variable delay time before probing the
spin noise. For long times the spins will reach thermal equilibrium,
where the noise of each atom in $F=4$ contributes $20/3$. The fraction
of atoms in $F=4$ is $9/16$, the remaining $7/16$ are in the $F=3$
state and do not contribute because of the large detuning.  Initially,
all atoms are in $F=4$ in the CSS and they each contribute the value 2
to the noise. Hence the measured noise must be on the form
\begin{equation}
  \text{Measured noise} \propto 2\cdot \exp(-\Gamma t) + \frac{20}{3}\cdot
   \frac{9}{16}(1-\exp(-\Gamma t)). 
\end{equation}
The predicted ratio of final to initial noise is thus $15/8\approx
1.88$. Experimentally we find the ratio $2.05\pm 0.09$ which is
consistent. To sum up, there is strong evidence that we really do
create the CSS with the correct minimum uncertainty noise. 

\subsubsection{Concluding Remarks on the Projection Noise Level}
\label{sec:conclude_projection}
Let us sum up the discussion of the projection noise level. To reach
the quantum noise limited performance one should first of all observe
the atomic noise grow linearly with the macroscopic spin size $J_x$. 
An experimental example of this was shown in
Fig.~\ref{fig:proj_noise}. The linearity of the noise basically arises
from the fact that different atoms yield independent 
measurements when their
spin state is detected. Technical noise sources from e.g.~external
electromagnetic fields couple to all atoms and the effect on the noise
variance would be quadratic. 

However, linearity alone is not enough. An ensemble of independent and
unpolarized atoms would also show a linear increase in the spin noise
when increasing the number of atoms. Since unpolarized atoms have
larger noise variance than the 100\% polarized atomic sample, we must
know independently that the spin orientation is high. In our
experiments the spin samples are polarized better than 99\%. 

One may argue that the small fraction of atoms that are not in the
completely polarized state could, in principle, form exotic-multi
particle states with a very high variance of the detected spin noise. 
The results discussed in Sec.~\ref{sec:thermal_spin_noise} prove that
this is not the case. 

Finally, as derived in App.~\ref{sec:atom_motionAll} the atomic motion
leads to an increased ratio of atomic to shot noise.  Generally a
large ratio of atomic projection noise to shot noise is good for the
quantum information protocols. However, we do not gain anything by the
increase of the atomic noise caused by atomic motion since there will
be an accompanying increase in the decoherence rate.

%
%
\subsection{Decoherence}
\label{sec:decoherence}

As mentioned earlier, all atoms are optically pumped into an extreme
Zeeman sublevel with the $x$-axis as quantization axis. A conventional
way of categorizing sources of decoherence is according to whether
they affect the magnitude of the spin along this axis or merely along
transverse directions. The appropriate life times of these are called
$T_1$ and $T_2$ and defined as:
\begin{equation}
  \label{eq:T1T2definition}
  J_x(t)=e^{-t/T_1}J_x(0)\qquad \mathrm{and}\qquad   
J_\mathrm{trans}(t)=e^{-t/T_2} J_\mathrm{trans}(0)
\end{equation}
in the absence of additional interactions. In the absence of the probe
light the dominant processes are collisions with the walls and other
atoms and, typically $T_1\approx300\mathrm{ms}$. Even though the
probe detunings are quite high in our experiments (700-1200MHz) making
the desired refractive Faraday interaction dominant by far, the small
probability of absorption still reduces $T_1$ by about a factor of 2
depending on probe power and detuning.  It is, however, still very
large compared to typical probe durations (0.5-2 ms).

The lifetime of the transverse spin components, however, turns out to
be much more critical to our experiments. As can be seen from
Eq.~(\ref{eq:spin_dens}) the transverse spin components are determined
by coherences between magnetic sublevels in the $x$-basis. Therefore
anything that affects the lifetime of $J_x$ will also affect $T_2$. In
addition, however, the total transverse spin components are also
sensitive to random phase changes in each atom.  As discussed in Sec. 
\ref{sec:bias_mag_field} we can use the widths obtained in MORS
signals in different experimental settings to quantify and separate
the effect of different decoherence mechanisms. The FWHM obtained from
such signals are related to $T_2$ by: 
\begin{equation}
  \label{eq:Gamma_transverse}
  \Gamma_{\mathrm{trans}}[\mathrm{Hz}]=\frac{1}{\pi T_2[\mathrm{s}]}
\end{equation}
\begin{figure}[t]
  \centering
  \begin{minipage}[t]{.48\textwidth}
  \centering
\includegraphics[width=1\textwidth,angle=0]{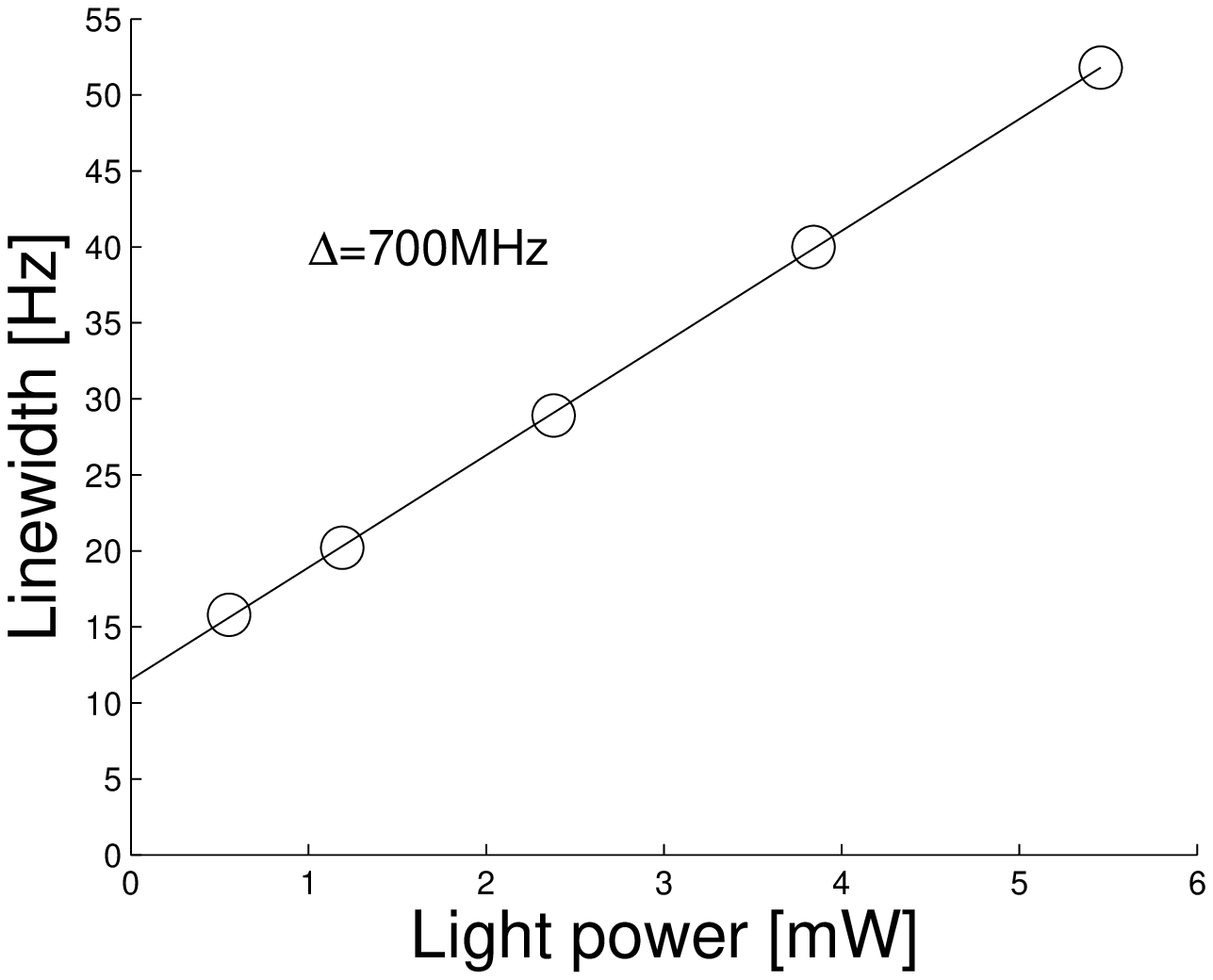}
\caption{\footnotesize  
  Linewidth of the $\sigma_{34}$ coherence in the MORS signal as a
  function of optical power. In a single trace it is impossible to
  separate the expected power broadening from absorption from the
  light induced collisions.  } \label{fig:powerbroadening}
\end{minipage}%
\hspace{0.5cm}%
  \begin{minipage}[t]{.48\textwidth}
    \centering
    \includegraphics[angle=0,width=\textwidth]{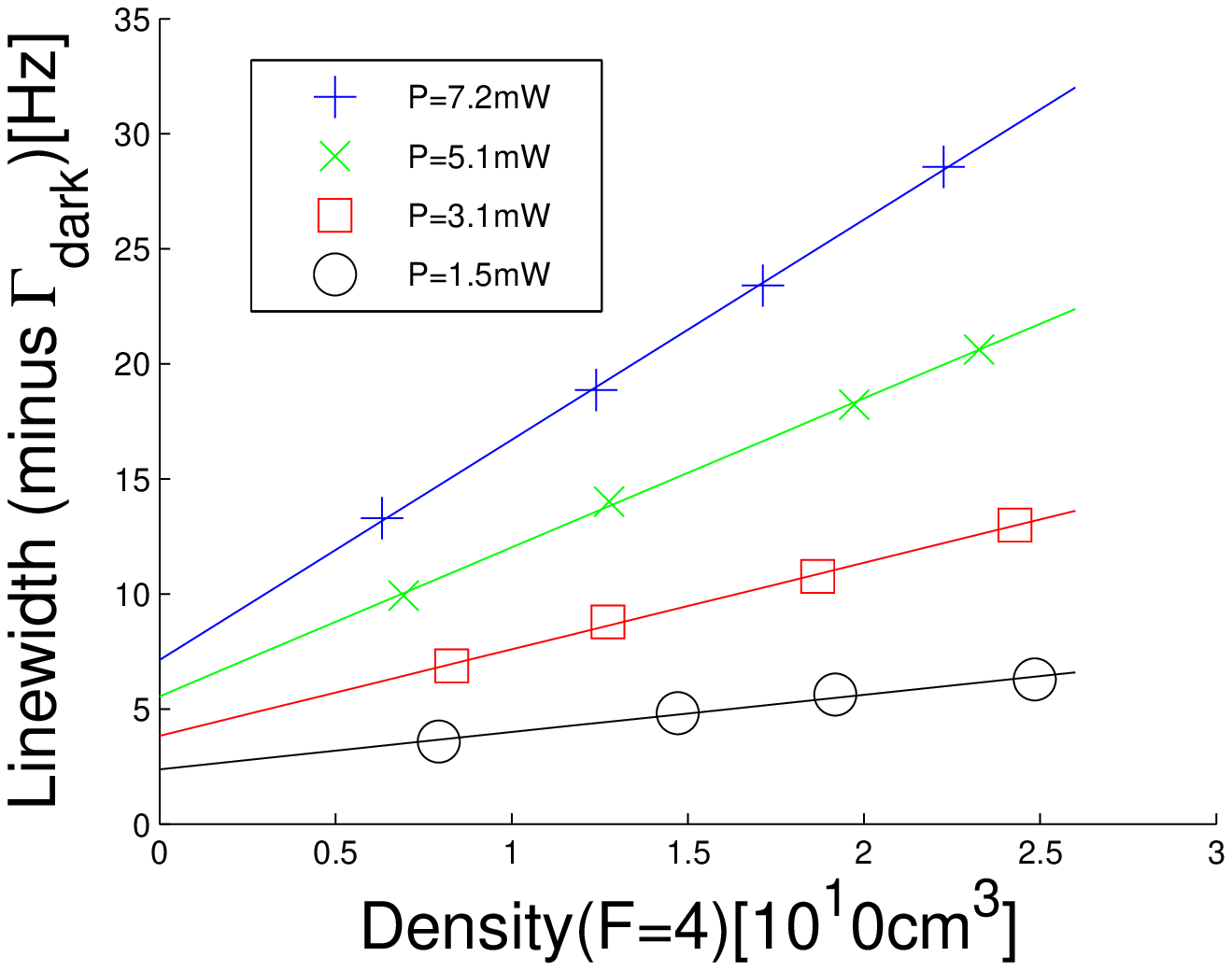}
    \caption{ \footnotesize  
      Linewidth of the $\sigma_{34}$ coherence in the MORS signal as a
      function of atomic density for different optical powers. The
      fact that the slopes are not equal reveals light induced
      collisions.}
  \label{fig:lightInducedDecoherence}
\end{minipage}
\end{figure}
We can separate the mechanisms into two main categories: some are
mediated by the probe light and the rest are independent of the
presence of the probe.  Starting with the latter, these combine to a
decay rate, $\Gamma_{\textrm{dark}}$, and consist mainly of phase
changing and spin-flip collisions with the walls and other atoms
and random phase changes because of atomic motion through
inhomogeneous magnetic fields. The effect of these are reduced by the
paraffin coating on the inside of the glass cells, the diluteness of
the atomic sample, and the application of additional dc-magnetic
fields to cancel field gradients. To determine
$\Gamma_{\textrm{dark}}$ we measure the width of the
$m_F=3\leftrightarrow 4$ coherence for different probe powers and find
the residual width in the absence of light. An example of this is
shown in Fig.~\ref{fig:powerbroadening}. As can be seen we obtain a
width of the order of 12 Hz corresponding to a lifetime of the
transverse spin of $T_2^{\textrm{dark}}\approx 27\mathrm{ms}$. We see
that this will limit but not destroy all correlations between two
subsequent pulses in e.g.  an entanglement experiment as described in
Sec.~\ref{sec:ent_gen_ver}.

Turning to the probe induced decoherence mechanism, we have already
mentioned absorption and subsequent spontaneous emission in the
discussion of $T_1$. Adding this effect to the other decoherence
mechanism, we would expect a total decoherence rate of the general
form:
\begin{equation}
  \label{eq:gamma_ideal}
  \Gamma_\mathrm{ideal}=a+b\cdot n+c\cdot P~,
\end{equation}
$n$ is the atomic density, $P$ is the light power, and $a$, $b$,
and $c$ are coefficients, which can be determined experimentally. If
$\Gamma$ is plotted vs. $n$ we would expect a line with constant slope
$b$ and offset determined by the optical power. In
Fig.~\ref{fig:lightInducedDecoherence} we show measurements of the
decoherence rate vs. atomic density for different optical powers. The
results clearly contradict the simple model of
Eq.~(\ref{eq:gamma_ideal}), since the slope grows with increasing
power. It turns out that the experiments fit a model:
\begin{equation}
  \label{eq:gamma_exp}
  \Gamma_\mathrm{exp}=a+b\cdot n+c\cdot P+d\cdot n\cdot P~,
\end{equation}
where the size of the expected pure power broadening term, $c$, agrees
with solutions of the Maxwell-Bloch equations for the full multi-level
atomic system in the presence of Doppler broadening. The last term
could represent light induced collisions, but a clear theoretical
understanding of the nature of these is still missing. For the
experimentally relevant densities and powers this term contributes
around 30 Hz of broadening and is thus the main source of decoherence.
We stress that this is a pure $T_2$ process since we do not observe
similar features when investigating the decay of the longitudinal
spin. Hence, the atoms practically decay towards the fully polarized
state, i.e. the coherent spin state. This was also assumed 
implicitly in the inclusion of decoherence for the entanglement and quantum
mapping protocols in Sec.~\ref{sec:protocols}, where decoherence was
modeled by an admixture of a vacuum state with the same variance as
the coherent spin state.

%
%
\subsection{Entanglement Results}
\label{sec:entanglement-results}

\subsubsection{Conditional Entanglement}
\label{subsec:condEnt}
\begin{figure}[t]
  \centering
  \begin{minipage}[t]{.48\textwidth}
  \centering
\includegraphics[width=1\textwidth,angle=0]{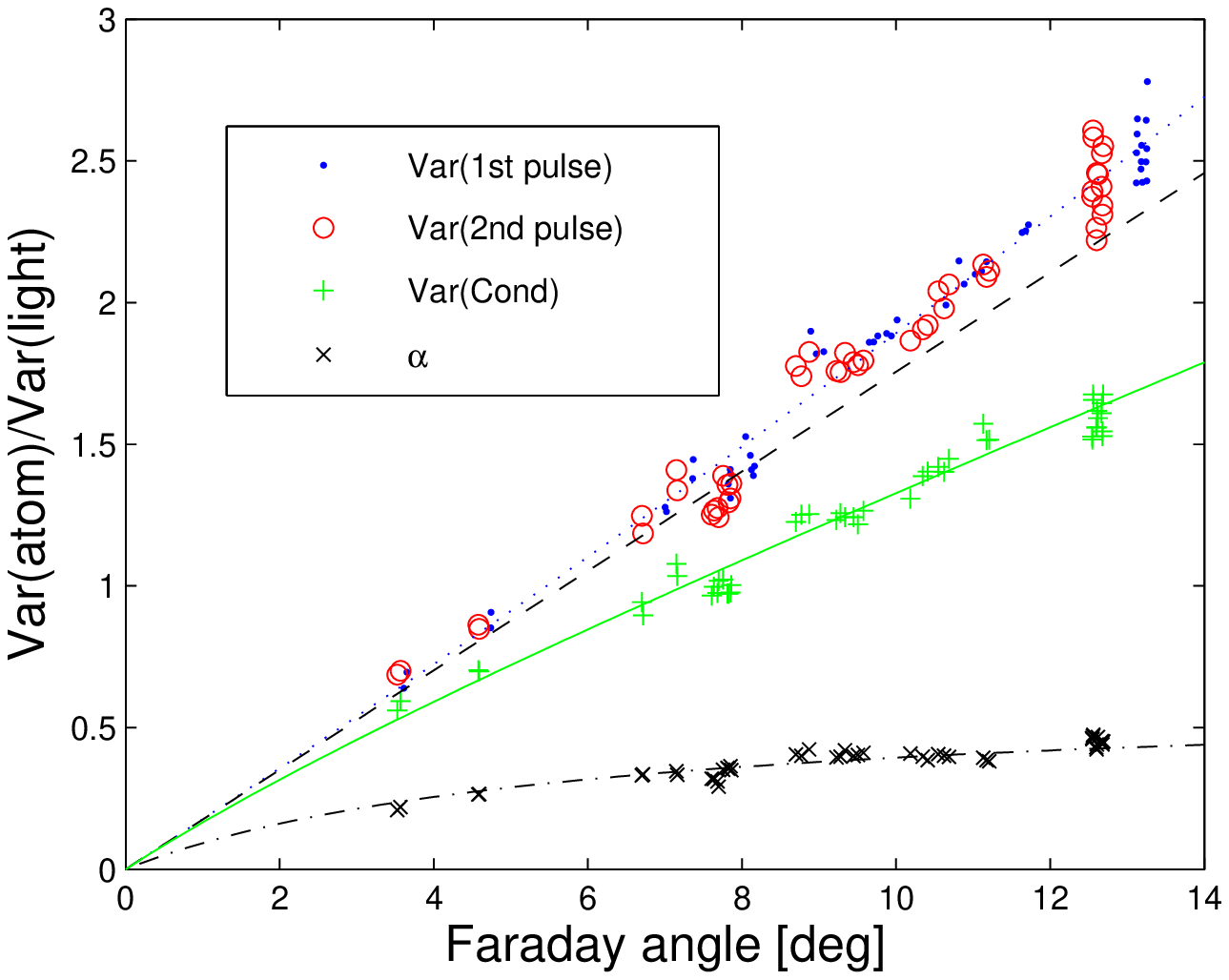}
\caption{\footnotesize  Atomic noise in units of shot noise as a
function of the macroscopic spin size (measured by DC Faraday
rotation). The dotted curve is a quadratic fit to the first pulse
variances (dots) and the dashed curve is the linear part of this.
Dash-dotted and full-drawn lines are fits to the the optimal weight
factors, $\alpha$, (x's) and the conditional variances (+'s)
respectively. } \label{fig:alldataentanglement}
\end{minipage}%
\hspace{0.5cm}%
  \begin{minipage}[t]{.48\textwidth}
    \centering
    \includegraphics[angle=0,width=\textwidth]{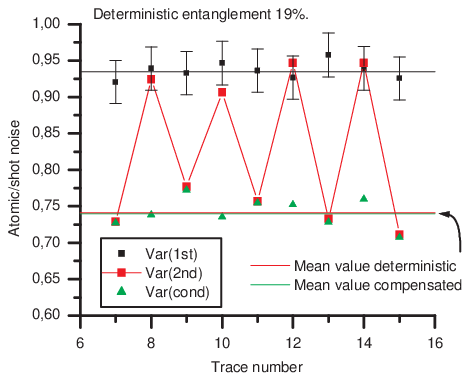}
    \caption{ \footnotesize  Deterministic entanglement generation.
    For a fixed atomic density negative feedback of the first pulse
    measurement is alternately turned on and off, thus switching
    between conditional and unconditional entanglement generation.
     }
  \label{fig:detent}
\end{minipage}
\end{figure}
We now turn to the experimental demonstration of entanglement
generation.  First the boundary between the classical and the quantum
fluctuations has to be established. As discussed in
Sec.~\ref{sec:proj-noise-level} this projection noise level is found
by performing several measurements of $\X{\mathrm{L}}$ as a function
of the macroscopic spin size and verifying a linear increase of the
atomic noise of each measurement. This linearity combined with nearly
perfect orientation of the sample ensures the correct projection noise
level. Once this is established we implement a probing sequence (see
Fig.~\ref{fig:exp_setup_typical}) in which the initial probing pulse
is followed by a second one after a short delay. To verify
entanglement we need to fulfill the
criterion~(\ref{eq:easy_critertion}), in which case our ability to
predict the outcome of the second probing of the atomic state
conditioned on the result of the first measurement exceeds the
classical limit.

For the experimental data we calculate the atomic part of the noise by
subtracting the shot and electronics noise of a single light pulse and
then normalize to the light shot noise level. In
Fig.~\ref{fig:alldataentanglement} the resulting atomic noise is shown
for the first and second pulses (with filled and empty circles
respectively). Since the measurement is of a QND-type and the variance
is calculated based on 10,000 independent repetitions of the pulse
sequence, the variance of the first and the second pulses should be
identical. We make a quadratic + linear fit to the first pulses'
variances (dotted curve) and from this extract the linear part, which
represent the coherent spin state level. The slope of this is
0.176(12), which can be compared to the theoretical values 0.187 and
0.165 obtained from Eq.~(\ref{eq:PN_theory_convenient}) with and
without the effect of atomic motion included. In this fit we used:
T=2.0ms, P=5.0mW, and $\Delta/2\pi =825$MHz. As discussed in
Sec.~\ref{sec:atom_motionAll} the results including atomic motion are
very preliminary, since this is an ongoing research topic. The overall
agreement with experimental results are, however, very encouraging.
Note that the small quadratic component is caused by various classical
noise sources.

Next we calculate the correlations between the first and the second
pulse measurements as discussed in Sec. \ref{sec:ent_gen_ver}. In
Fig.~\ref{fig:alldataentanglement} the pulses show the conditional
variance $\var(A_2|A_1)+\var(B_2|B_1)$ normalized to shot noise and
with shot and electronics noise subtracted. The points below the
straight line~(\ref{eq:easy_critertion}) indicate that we have in fact
created an entangled state between the two atomic samples. For the
highest densities the noise reduction is up to 30\%.  The
corresponding $\alpha$-parameters from the minimization
procedure~(\ref{eq:alpha}) are plotted in
Fig.~\ref{fig:alldataentanglement} with crosses.  Ideally we would
expect these to follow Eq.~(\ref{eq:alpha_theory}) but since the atoms
decohere the appropriate expression is instead
Eq.~(\ref{eq:alpha_theory_decoh}). Fitting, we obtain
$\beta=0.619(11)$, which is inserted into
Eq.~(\ref{eq:var_theory_decoh}) (solid curve). Considering the fact
that the solid and dash-dotted curves are obtained from a single free
parameter, $\beta$, the agreement between experimental results and the
simple model of decoherence must be considered very satisfactory. In
addition, we would like to stress the very important point that the
two atomic samples, which are entangled, are in completely separate
environments about 0.5 meters apart \citep{sherson:leshouche}. This
represents a major breakthrough towards the creation of truly distant
entanglement, which combined with quantum teleportation will enable
quantum communication over long distances.
\subsubsection{Unconditional Entanglement}
\label{subsec:uncondEnt}
As we have just seen, the results of two probes of the spin state yield
correlated results. The actual results, however, vary from shot to
shot, representing random realizations of the probability distribution
of the spins. That is, we create a non-local state with reduced
variance but with a non-deterministic mean value. Thus, the
entanglement only appears when the knowledge gained in the first pulse
is applied. To create a deterministically entangled state in which no
knowledge of measurement results is necessary would of course
constitute a very important advance. We have realized this
experimentally by simply feeding the result of the first measurement
pulse back to the atoms using an RF-magnetic pulse as discussed in
Sec.~\ref{sec:manipulating_spin_state}. This procedure is very closely
related to the way in which unconditional spin squeezing is generated
in~\citep{geremia:04} except that there the feedback is applied
continuously in time, which is more robust against errors in the
feedback strength.

In the experiment we first choose a certain atomic density. Using the
linear fit of Fig.~\ref{fig:alldataentanglement} we find the
projection noise level relative to which entanglement is estimated.
Next, we vary the strength or the phase of the RF feedback pulse, and
observe the variance of the second pulse fluctuations as a function of
these feedback parameters. The feedback giving the lowest variance
corresponds to the feedback with the optimal gain. In
Fig.~\ref{fig:detent} we alternate between having feedback and not
having it. As can be seen clearly, with feedback we obtain the same
variance for the second pulse as we do without feedback for the
conditional variance. This means that we have in fact created a state,
which is deterministically entangled. That is, the state after the
first probe and subsequent feedback of the measurement result has zero
mean and a variance reduced by 19\% compared to the projection noise
level. When feedback is turned on the variance of the second pulse and
the conditional variance coincide, which means that knowledge of the
measurement result cannot improve the degree of entanglement beyond
the entanglement proved by the second pulse measurement alone.

%
%
\subsection{Quantum Memory Results}
\label{sec:quant-map-results}
\begin{figure}[t]
\centering
  \begin{minipage}[t]{.48\textwidth}
\centering
 \includegraphics[width=\textwidth]{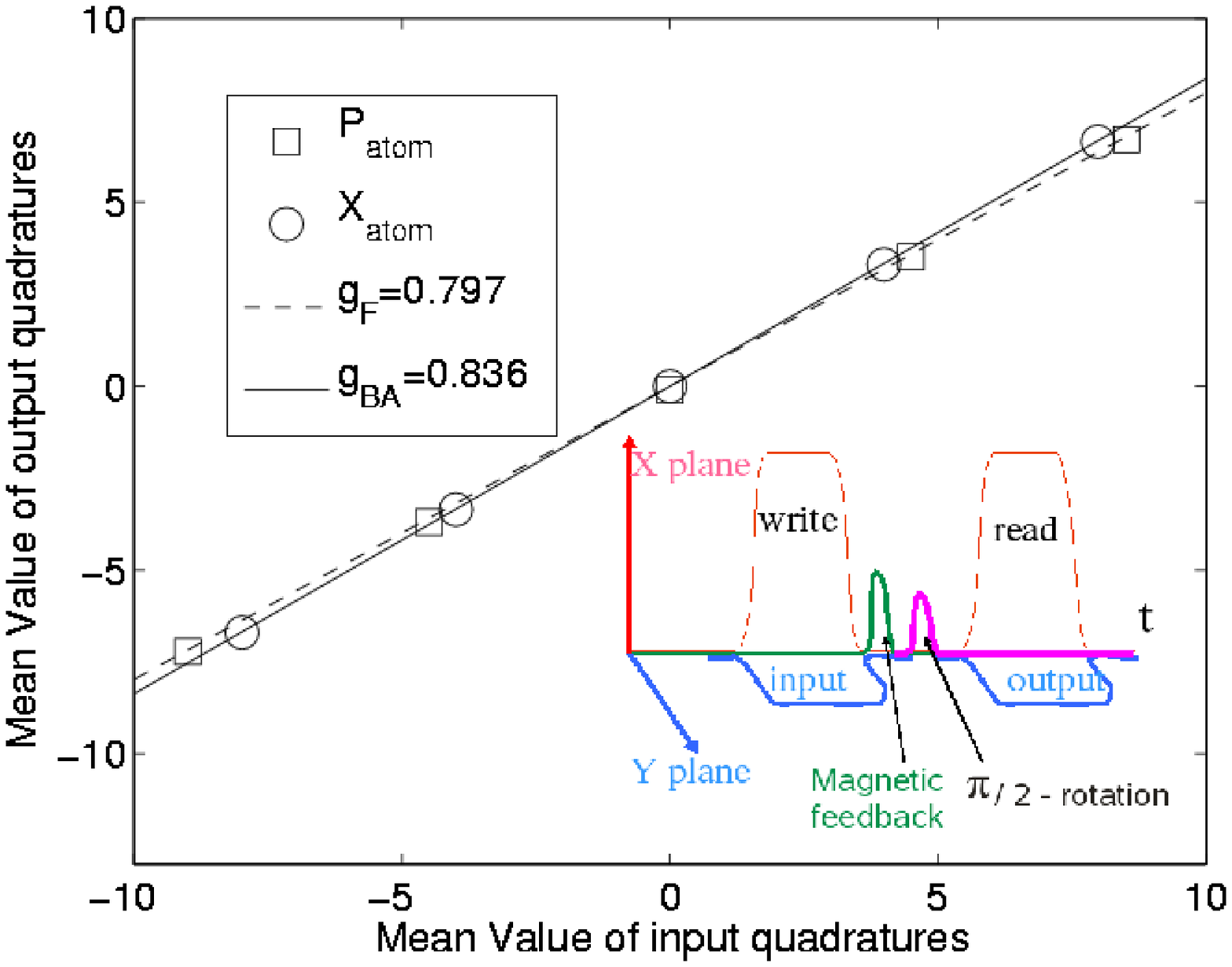}
 \caption{\small Mean value of the read out pulse as a function of the
   mean values of the input light variables,
   $\X{\mathrm{L}}^{\mathrm{in}}$ and $\P{\mathrm{L}}^{\mathrm{in}}$,
   to be stored. Inset: the strong classical and weak quantum pulses
   in opposite polarizations.  Between the input and the output pulses
   are the feedback pulse and the optional $\pi/2$ pulse.  }
 \label{fig:memorymean}
 \end{minipage}%
\hspace{0.5cm}%
  \begin{minipage}[t]{.48\textwidth}
\centering
 \includegraphics[width=\textwidth]{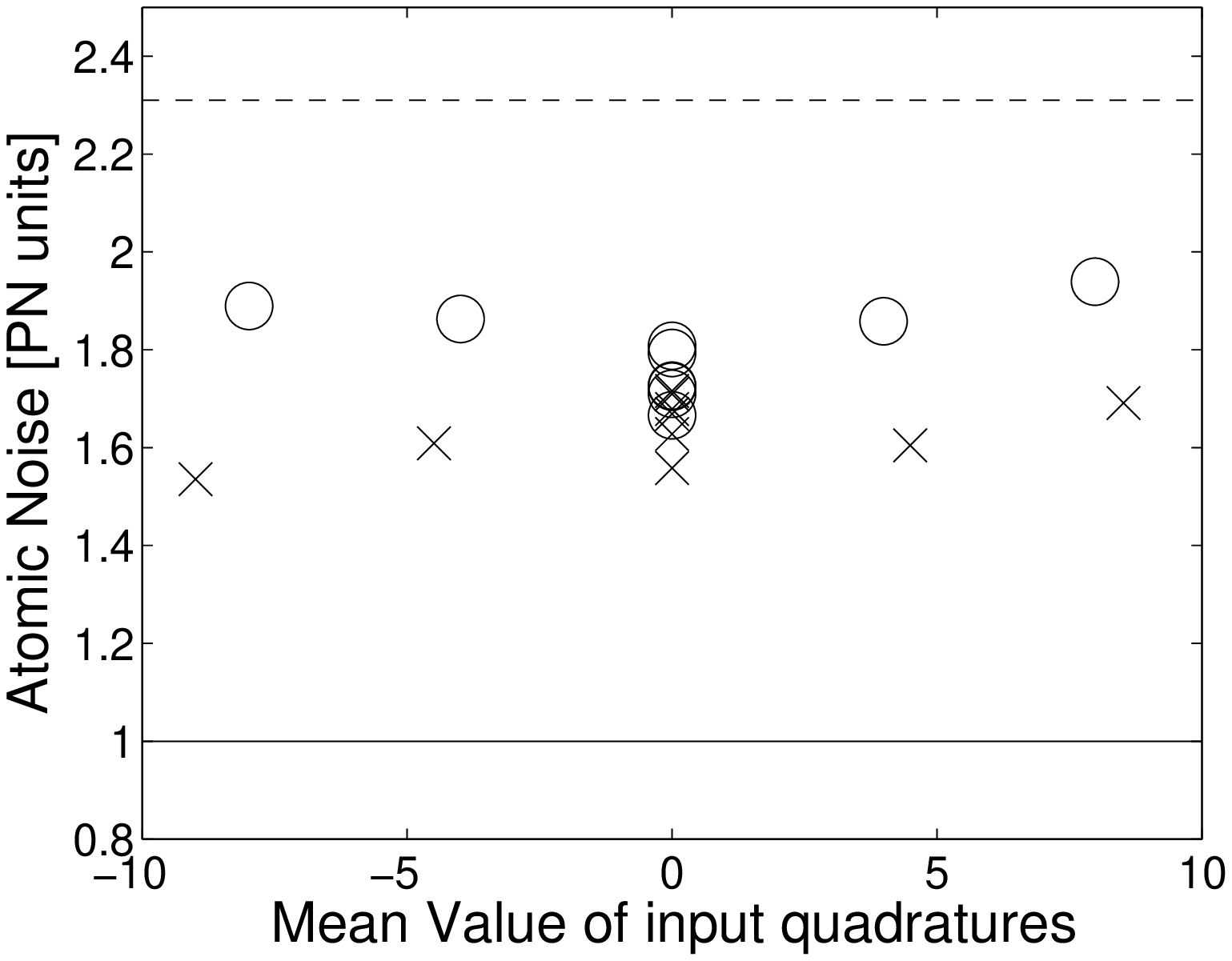}
 \caption{\small Variance of the atomic quadratures,
   $\X{\mathrm{A}}^{\mathrm{out}}$(o's) and
   $\P{\mathrm{A}}^{\mathrm{out}}$(x's), as a function of the mean
   values of the input light variables. Full drawn curve: the variance
   for perfect quantum storage. Dotted curve: classical limit on the
   variances for $n_0=4$}
 \label{fig:memoryvar}
\end{minipage}
\end{figure}

We will now present the experimental demonstration of quantum memory,
which has been published in~\citep{ourNature04}. The used protocol was
discussed in Sec.~\ref{sec:quantum-memory-protocols} and, briefly,
involves 1) preparation of the initial atomic state in a coherent
state via optical pumping, 2) mapping of one of the light quadratures
through the off-resonant Faraday interaction,
Eqs.~\ref{interact_light}-d, and 3) mapping of the second light
quadrature by a direct measurement and subsequent feedback (an
illustration of the timing sequence is shown in the inset of
Fig.~\ref{fig:memorymean}). For the reasons discussed in
Sec.~\ref{sec:retrieval} we have not been able to retrieve the mapped
state. Instead we have performed a destructive reconstruction of the
mapped state. This is done by waiting for a time $\tau$ and then
sending a readout light pulse through the atomic sample. Measuring the
$\X{}$-component of the outgoing light will then give information
about the stored atomic $\P{}$-component according to
Eq.~(\ref{interact_light}). If a $\pi/2$ rotation in the atomic $XP$
space is performed prior to the readout pulse we obtain information
about the stored atomic $\X{}$-component. Repeating this 10,000 times,
the statistics for the atomic variables after the storage procedure
can be reconstructed. The first thing to check is that the mean value
of the stored state depends linearly on the mean value of the input
light state. This is shown in Fig.  \ref{fig:memorymean}. First of all
we note that the linear dependence is clear for both quadratures. This
completes the proof of classical memory performance. The next thing to
note is that the slope is not unity, which means that the stored state
has a different mean value than the input state. The reason for this
is discussed further below.  For the quadrature mapped straight
from the back action of the light onto the atoms we have the gain of
$g_{\mathrm{BA}}=0.836$ and for the quadrature mapped via the feedback
we have $g_{\mathrm{F}}=0.797$.
In order to verify quantum storage we also need to consider the
shot-to-shot fluctuations in the stored state, which for a Gaussian
state are fully characterized by the variance of the state. The
experimentally reconstructed variances of the atomic quadratures,
that is the variance of the readout pulse with the one unit of shot
noise intrinsic to the read-out pulse subtracted, are shown in Fig.
\ref{fig:memoryvar}. Also shown is the ideal quantum limit on the
variance for a perfect mapping and the classical limit for $n_0=4$.
As can be seen the variance is more or less independent of the mean
value of the input light quadratures. The fidelity of the stored
state for a Gaussian distribution of input states with mean photon
number $n_0$ can be calculated given the measured gains and
variances, $\sigma_x$ and $\sigma_p$, according to:
\begin{equation}
  \label{eq:qmfidel}
  F = \frac{2}{\sqrt{(2n_0(1-g_{\mathrm{BA}})^2 + 1 + \sigma_x)(2n_0(1-g_{\mathrm{F}})^2
      +1+\sigma_p)}}~.
\end{equation}
With the experimentally measured values we get $F=(66.7\pm 1.7)\%$
for $n_0=4$ and $F=(70.0\pm 2.0)\%$ for $n_0=2$. The best classical
fidelity was recently derived by~\citet{hammerer:05}
\begin{equation}
  \label{eq:ClassFidel}
  F_{\mathrm{class}}=\frac{1+\bar{n}}{1+2\bar{n}}\to \frac{1}{2}~,
  \qquad \bar{n}\to \infty
\end{equation}
for coherent states drawn out of a Gaussian distribution with mean
photon number $\bar{n}$. This means that $F_{\mathrm{class}}$
decreases monotonically from unity for the vacuum state to $1/2$ for an
arbitrary coherent state. For $\bar{n}=4,2$ we obtain the classical
boundaries of 55.6\% and 60.0\% respectively. This verifies that the
storage of the light state in fact constitutes a quantum mapping. The
results shown were obtained for a pulse duration of 1 ms and the
memory has been shown to work for up to 4 ms delay between the two
probe pulses. Note that we have chosen to calculate the fidelity as
the average of the squared overlap between the stored state and the
ideally stored state.  For non-unity gain this decreases very rapidly
with coherent states having large amplitudes. However, one could argue
that a storage with an arbitrary but known gain constitutes just as
useful a memory as unity gain memory. If analyzed solely in terms of
the added noise, our memory would therefore perform much better than
the previously stated results, which can therefore be viewed as a
lower bound on the memory capability.

Note also that the choice of the optimal gain which maximizes the
fidelity depends on the class of available states. For classes of
coherent states limited in their amplitudes discussed in this section,
the experimental gains quoted above are, in fact, close to optimal
ones.

\subsubsection{Decoherence}
\begin{figure}[t]
 \centering
  \includegraphics[width=0.5\textwidth]{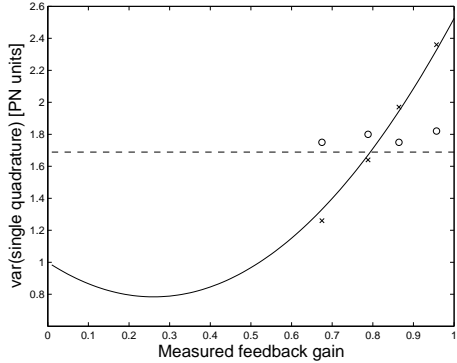}
  \caption{\small Reconstructed variances of
    $\X{\mathrm{A}}^{\mathrm{out}}$(o's) and
    $\P{\mathrm{A}}^{\mathrm{out}}$(x's) as a function of feedback gain.
    The curves are theoretical curves based on independently
    determined noise parameters.}
  \label{fig:feedback}
\end{figure}
In Fig. \ref{fig:feedback} we show the final atomic variances as a
function of the feedback gain. These are compared to the variances
expected from Eqs.~(\ref{eq:vardecoher}) with independently determined
decoherence values of $\beta=0.61$ and $\zeta=0.75$. We stress that
this is not a fit to the data, which means that we understand the
level of experimentally determined atomic noise quite well. The figure
clearly shows that, because of the decoherence and light loss, if the
feedback gain is increased towards unity the noise grows
dramatically. With this, the fidelity quantified by Eq.
(\ref{eq:qmfidel}) can be optimized with respect to the feedback gain.
As can be seen from the values of $\beta$ and $\zeta$ the light loss
and atomic decoherence are significant. The high light loss is due to
the fact that the glass cells containing the atomic vapor were not
anti-reflection coated. Therefore each glass-air interface contributes
about 4\% loss. The main source of atomic decoherence is light
assisted collisions, which change the phase of the atomic coherence
without affecting the macroscopic spin size $J_x$. The atoms are
driven towards a coherent state, which justifies the use of the
simple model of beam splitter admixture of vacuum.

%
%
\section{Conclusions}
We have described the progress in development of the deterministic
quantum interface between light and macroscopic atomic ensembles.
Theory of the interface is based on canonical variables which provide
a convenient language when different physical systems, such as light
and atoms, are considered.  Surprisingly enough, simple dipole
off-resonant interaction of light with spin polarized atomic samples
with high optical density provides a powerful tool for quantum state
engineering and transfer. To perform sophisticated quantum information
protocols, the interaction can be combined with a quantum measurement
on light and subsequent feedback to the atoms. Two central
experiments of this review, entanglement of two macroscopic objects
and quantum memory for light, are performed following these general
steps.

Future perspectives include multipass protocols towards more efficient
quantum memory including the retrieval process, as well as various
types of quantum teleportation involving atomic ensembles. Another
exciting future direction is demonstration of deterministic qubit
state manipulation with the tool box described in this paper. It is
appropriate to stress that the language of canonical variables is
fully applicable not only to Gaussian states but to all single mode
states, including qubit and Fock states. This language is associated
with homodyne measurements in quantum optics. Addition of single
photon counting to the experiments described in this article may pave
the road towards more efficient and robust quantum information
processing and communication.

\section*{Acknowledgements}
This research has been supported by the Danish National Research
Foundation and by EU grants QUICOV and COVAQIAL. B.~Julsgaard is
supported by the Carlsberg Foundation. We would like to thank J.I.
Cirac, L.M.Duan, K. Hammerer, J. Fiurasek, A. Kozhekin, D. Kupriyanov,
A. Kuzmich, and K. M\o lmer for fruitful collaborations. The
contributions of J. Hald and J.L. S\o rensen to the early experiments
are gratefully acknowledged.

\bibliographystyle{my_elsart-harv}
\bibliography{bibfile}

\appendix


\section{Effect of Atomic Motion}
\label{sec:atom_motionAll}
Since in our experiments the atoms are at room temperature and, for
experimental reasons, the light beam does not cover the entire cross
section of the atomic sample, the atoms move across the beam
several times ($\sim 10$) during the time of a pulse. This averaging
effect insures that all atoms spend roughly the same amount of
time inside the beam but, as we shall see, it still has important
implications for the noise properties. In brief, the atomic motion
modifies the projection noise level and acts as an additional source
of decoherence since two subsequent probe pulses interact with
the atoms differently. The results are related to the work
of~\citet{kuzmich:04}.

\subsection{Modeling Atomic Motion}
\label{sec:atom_motion}
To be more quantitative, we introduce new pseudo-angular momentum
operators $J_q\to \sum_i p_i F_q^{(i)}$, where $p_i$ is the fraction
of time the $i$'th atom spends interacting with the laser beam and
$q=x,y,z$.  These have the commutator:
\begin{equation}
  \left[\sum_{i=1}^{N_{\mathrm{at}}} p_i F_y^{(i)},
    \sum_{i=1}^{N_{\mathrm{at}}} p_i F_z^{(i)}\right]
  = \sum_{i=1}^{N_{\mathrm{at}}} p_i^2\left[F_y^{(i)},F_z^{(i)}\right]
  = i\sum_{i=1}^{N_{\mathrm{at}}} p_i^2 F_x^{(i)}. 
\end{equation}
This leads to the Heisenberg uncertainty relation (for a highly
polarized sample with $F_x \approx F$)
\begin{equation}
  \var\left(\sum_{i=1}^{N_{\mathrm{at}}} p_i F_y^{(i)}\right)
  \var\left(\sum_{i=1}^{N_{\mathrm{at}}} p_i F_z^{(i)}\right)
    \ge \left(\frac{J}{2}p^2(1+\sigma^2)\right)^2
\end{equation}
where we have introduced the mean $p = \left<p_i\right>$ and
variance $\var(p_i) = \sigma^2\cdot p^2$ of $p_i$. With this
definition, $\sigma$ is the relative standard deviation of $p$. Since, 
for the coherent spin state, 
\begin{equation}
\label{eq:var(CSS)}
  \var\left(\sum_{i=1}^{N_{\mathrm{at}}} p_i F_z^{(i)}\right) = \frac{J}{2}p^2(1+\sigma^2)
   = \var(\mathrm{CSS}),
\end{equation}
this highly polarized state corresponds to a minimum
uncertainty state. The measured noise is then limited by the
Heisenberg uncertainty principle and we confidently call this
projection noise. To maintain the correct commutation relation $[X,P]
= i$ we experimentally normalize the atomic operators to the
\emph{measured} projection noise, i.e.~instead of defining $X =
J_y/\sqrt{J}$ we effectively have $X = \sum_i p_i
F_y^{(i)}/\sqrt{Jp^2(1+\sigma^2)}$.

The average fraction of time $p$ each atom spends inside the beam is
clearly $p = A_{\mathrm{beam}}/A_{\mathrm{cell}}$.  Let us now
discuss the scaling of $\sigma^2$ with simple physical parameters. 
The fact that the variance may be non-zero arises from the finite
time available for the averaging process carried out by the atomic
motion. A typical traversing time across the vapor cell is $\tau =
L/v_0$ where $L$ is the cell dimension and $v_0$ is e.g.~the
one-dimensional rms speed of the atoms. We may think of this atomic
motion as $n$ independent journeys across the vapor cell volume,
where $n \approx T/\tau = Tv_0/L$. We then model the motion through
the beam with mean occupancy $p$ by assuming in each walk across the
cell volume that either (1) the atom spends all the time $\tau$
inside the beam, which should happen with probability $p$, or (2) the
atom spends all the time $\tau$ outside the beam which should happen
with probability $1-p$. 

We then count the number of times $n_{\mathrm{inside}}$ that an atom
was inside the beam out of the possible $n$ journeys. In this simple
model $n_{\mathrm{inside}}$ is a stochastic variable which is
binomially distributed with mean $np$ and variance $np(1-p)$. We are
interested in the fraction of time ($\approx n_{\mathrm{inside}}/n$)
spent inside the beam. It follows
$\left<n_{\mathrm{inside}}/n\right> = p$ and $\sigma^2 =
\var([n_{\mathrm{inside}}/n]/p) = (1-p)/np$. Hence the simple model
leads to
\begin{equation}
  \label{eq:simple_model_mean_var}
  p = \frac{A_{\mathrm{beam}}}{A_{\mathrm{cell}}}
  \quad\text{and}\quad
  \sigma^2 = \frac{(A_{\mathrm{cell}} - A_{\mathrm{beam}})L}
{A_{\mathrm{beam}} T v_0}
\end{equation}
Note the characteristic scaling with $T^{-1}$ and with the area
$(A_{\mathrm{cell}} - A_{\mathrm{beam}})$ \emph{not} covered by the
light beam (when $A_{\mathrm{beam}}$ is close to its maximum value
$A_{\mathrm{cell}}$). We note that due to the simplicity of the
above model the absolute numbers should only hold as an order of
magnitude estimate. Numerical simulations performed for a cubic cell
have shown that the relative variance $\sigma^2$ is roughly four
times smaller than the estimate above.  Also, due to the Doppler
broadening, the effective detuning differs from atom to atom and
causes an increase
in $\sigma^2$.  

\subsection{Atomic Motion as a Source of Decoherence}
To see how atomic motion acts as an effective source of decoherence,
imagine that we perform some manipulations of atoms using one laser
pulse and subsequently probe these manipulations with another laser
pulse. Since atoms move during interactions the probed quantum
operator changes in time.  Comparing the operator at the 1st and 2nd
times we get
\begin{equation}
\label{eq:decoh_motion}
\begin{split}
  &\var\left(\sum_{i=1}^{N_{\mathrm{at}}} p_{i,\mathrm{2nd}}
    F_{z,\mathrm{2nd}}^{(i)} -\sum_{i=1}^{N_{\mathrm{at}}}
    p_{i,\mathrm{1st}} F_{z,\mathrm{1st}}^{(i)}\right) =
  \sum_{i=1}^{N_{\mathrm{at}}} \var(F_{z,\mathrm{1st}}^{(i)} )
  \left<(p_{i,\mathrm{2nd}}-p_{i,\mathrm{1st}})^2\right> \\
  &\quad = \frac{J}{2}\cdot 2 p^2\sigma^2 =
  2\var(\mathrm{CSS})(1-\beta) \quad\text{with}\quad \beta =
  \frac{1}{1+\sigma^2}.
\end{split}
\end{equation}
We assumed $p_{i,\mathrm{1st}}$ and $p_{i,\mathrm{2nd}}$ to be
uncorrelated, which is reasonable since a collision with the cell
wall randomizes the velocity of the atoms. Also, we took
$F_{z,\mathrm{1st}}^{(i)} = F_{z,\mathrm{2nd}}^{(i)}$. This
corresponds to having no decoherence at all apart from the effect of
atomic motion which is the only effect studied in this calculation. 
To interpret the above calculations we consider a standard
decoherence calculation.  Consider a true spin operator $J_z$
subject to decoherence parametrized by the number $\beta$ such that
\begin{equation}
  J_{z,\mathrm{1st}} \rightarrow J_{z,\mathrm{2nd}}
    = \beta J_{z,\mathrm{1st}} + \sqrt{1-\beta^2}J_{\mathrm{vac}}
  \quad\text{with}\quad
  \var(J_{\mathrm{vac}}) = \frac{J}{2} = \var(\mathrm{CSS}). 
\end{equation}
Then the operator changes by an amount characterized by the variance
\begin{equation}
\begin{split}
  \var\left(J_z^{\mathrm{2nd}} - J_z^{\mathrm{1st}}\right)
  &= \var\left(J_z^{\mathrm{1st}}(1-\beta)
              -\sqrt{1-\beta^2}J_{\mathrm{vac}}\right) \\
  &= J(1-\beta) = 2\var(\mathrm{CSS})(1-\beta)
\end{split}
\end{equation}
which is exactly the same as in~(\ref{eq:decoh_motion}). We
are led to the conclusion that atomic motion inevitably gives
rise to an effective decoherence. We thus see that, whereas the
increased coherent spin state noise with increased $\sigma^2$ might
seem to suggest that non-classical states are more easily created (by
producing states with noise lower than the CSS), this is compensated
for by an increased decoherence of the state. Therefore, higher
$\sigma^2$ does not lead to higher fidelity protocols.

\section{Technical Details}
\label{sec:tech-details}
\subsection{Light Polarization and Stark Shifts}
\label{sec:polarization_and_Stark}
Let us calculate the Stark effect from the probe laser on the magnetic
sublevels $\ket{F,m}$. We let the light be strong and linearly
polarized along the vector
\begin{equation}
  \vec{e}_1 = \vec{e}_x\cos\alpha + \vec{e}_y\sin\alpha,
\end{equation}
i.e.~$\alpha$ is the angle between the macroscopic spin direction (the
$x$-axis) and the probe polarization direction. Light is propagating
in the $z$-direction. The Stark effect on magnetic sub-levels is much
weaker than the splitting caused by the constant bias magnetic field
and can be calculated in non-degenerate perturbation theory from the
interaction Hamiltonian~(\ref{sec_inter:final_eff_Hamil_real}). The
$a_0$ term is common to all levels, the $a_1$ term is zero on average
since $\mean{\S{z}} = 0$, and we are left with the higher order
components proportional to $a_2$. 

\begin{figure}[t]
    \centering
    \includegraphics[angle=0,width=0.5\textwidth]{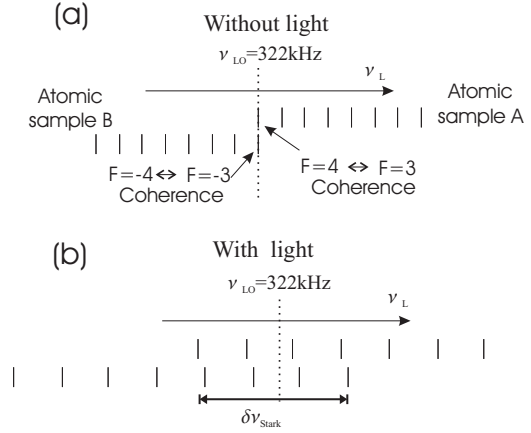}
    \caption{ \footnotesize  An illustration of the problem with the
      light induced Stark shift of the Zeeman sublevels. Without
      applying additional fields the two important Zeeman resonances
      cannot be overlapped both without light (a) and with light (b).} 
  \label{fig:stark_shift}
\end{figure}

With $\a{1}$ being the creation operator along the strong direction we
have $\a{x} = \a{1}\cos\alpha$ and $\a{y} = \a{1}\sin\alpha$
(neglecting the orthogonal direction to $\a{1}$ which will be in the
vacuum state for linear polarization). With $\S{\pm} = \S{x}\pm
i\S{y}$ we derive from~(\ref{def:intro_stokes}) that
\begin{equation}
  \mean{\S{+}(t)} = \frac{\phi(t)}{2}e^{2i\alpha}
  \quad\text{and}\quad
  \mean{\S{-}(t)} = \frac{\phi(t)}{2}e^{-2i\alpha},
\end{equation}
where $\phi(t)$ is the photon flux and Stokes operators are normalized
to photons per second. In order to calculate the higher order terms of
the interaction Hamiltonian for a single atom we leave out the
integral $\int \ldots \rho A dz$
in~(\ref{sec_inter:final_eff_Hamil_real}) and renormalize light
operators (by absorbing the speed of light $c$) to photons per second
and find
\begin{equation}
\label{eq:Stark_Hamiltonian}
\begin{split}
\H_{\mathrm{int}}^{\mathrm{eff}} = &-\frac{\hbar \gamma}{8 A \Delta}
\frac{\lambda^2}{2\pi}a_2\cdot\phi(t)\cdot\left( \j{z}^2 -
  [\j{x}^2-\j{y}^2]\cos(2\alpha)-[\j{x}\j{y}+\j{y}\j{x}]\sin(2\alpha)\right). 
\end{split}
\end{equation}
We also replaced $\j{\pm}$ by $\j{x}\pm i\j{y}$. We need to calculate
the expectation value of this Hamiltonian for the different energy
eigenstates $\ket{m}$ quantized along the $x$-direction. We have
\begin{equation}
  \begin{split}
    \bra{m}\j{x}^2\ket{m} &= m^2, \\
    \bra{m}\j{y}^2\ket{m} &= \frac{F(F+1)-m^2}{2}, \\
    \bra{m}\j{z}^2\ket{m} &= \frac{F(F+1)-m^2}{2}, \\
    \bra{m}\j{x}\j{y}+\j{y}\j{x}\ket{m} &= 0. 
  \end{split}
\end{equation}
The first of these is obvious since $\ket{m}$ is quantized along the
$x$-axis. We have $\mean{\j{y}^2} = \mean{\j{z}^2}$ by symmetry and
the value is found from the fact that $\j{x}^2 + \j{y}^2 + \j{z}^2 =
F(F+1)$. Also, by symmetry we have in an eigenstate of $\j{x}$ that
$\bra{m}\j{y}\j{x}\ket{m} = m\cdot\bra{m}\j{y}\ket{m}= 0$ which leads
to the final line. Calculating the expectation value
of~(\ref{eq:Stark_Hamiltonian}) we get
\begin{equation}
  E_m^{\mathrm{Stark}} = \frac{\hbar\gamma}{8A\Delta}\frac{\lambda^2}{2\pi}
    a_2\cdot\phi(t)\cdot\left[\frac{1+3\cos(2\alpha)}{2}\cdot m^2
    - \frac{1+\cos(2\alpha)}{2}F(F+1)\right]. 
\end{equation}
What is important for us is the shift $\delta\nu_{m+1,m} =
(E_{m+1}^{\mathrm{Stark}}-E_m^{\mathrm{Stark}})/h$ of a magnetic
resonance line which then becomes
\begin{equation}
  \delta\nu_{m+1,m}^{\mathrm{Stark}}[\mathrm{Hz}]
   = \frac{\gamma\lambda^2 a_2}{64\pi^2 A\Delta}
     \cdot\phi(t)\cdot[1+3\cos(2\alpha)]\cdot[2m+1]. 
\end{equation}
This Stark shift is problematic for several protocols, especially the
setup with two oppositely oriented samples. Note, that for atoms
polarized in the $m_F = 4$ state the relevant transition $m_F = 4
\leftrightarrow m_F = 3$ has a Stark shift proportional to $2\cdot 3 +
1 = 7$. An oppositely oriented sample with $m_F = -4$ has for the
transition $m_F = -3 \leftrightarrow m_F = -4$ a Stark shift
proportional to $2\cdot(-4)+1 = -7$. Hence, these two transition
frequencies cannot be overlapped both in the presence and absence of light
(see Fig.~\ref{fig:stark_shift} for an illustration).

One remedy for this is to choose the light to be linearly polarized
at an angle $\alpha = 54.7^{\circ}$ corresponding to
$3\cos(2\alpha) = -1$, and the Stark term disappears. As we shall see
in Sec.~\ref{sec:laser_noise} this gives rise to other problems. 

Another remedy is to add an extra bias magnetic field along the
$x$-direction when the laser light is on. In this way the frequency of
the desired transitions can be kept stable. This is the approach we
have taken and it works well. One should note though, that with our
laser pulse timing in the vicinity of one millisecond, it is not
completely trivial to make a magnetic pulse following the laser
intensity since eddy currents in metallic parts and induced electric
fields in other current loops for magnetic fields should be taken into
account. A convenient diagnostic method is to apply a classical shift
to the spin states along the lines of
Eq.~(\ref{eq:RF-field-influence}) prior to the application of two
laser pulses. An $\S{y}$ detection of the light will show in real time
the mean value of the spin state components $\J{y}'$ and $\J{z}'$.
These should be constant through each laser pulse and conserved in the
dark time in between (apart from a possible decay) if all frequencies
are well adjusted.

\subsection{Influence of Laser Noise}
\label{sec:laser_noise}
In this section we make a brief comment on laser noise entering 
the atomic samples. This is by no means a complete analysis but
it describes some important issues in connection with the choice of
laser polarization. We start out with the interaction
Hamiltonian~(\ref{sec_inter:final_eff_Hamil_real}) and also assume
that the atoms are polarized along the $x$-axis and placed in a bias
magnetic field, leading to the magnetic Hamiltonian $\H_{\mathrm{mag}}
= \Omega \j{x}$. This leads to the following equations of motion for
the transverse spin components $\j{y}$ and $\j{z}$:
\begin{align}
\label{eq:propagation_jy}
  \frac{\partial}{\partial t} \j{y}(z,t) &= -\Omega \j{z} +
   \frac{c\gamma}{8A\Delta}\frac{\lambda^2}{2\pi}
    \left\{-a_1\S{z}\j{x} \right . \\
   \notag
   &\qquad + a_2 \left. \left(-(2\S{x}+\hat{\phi})[\j{x}\j{z}+\j{z}\j{x}]
       -2\S{y}[\j{z}\j{y}+\j{y}\j{z}]  \right) \right\}, \\
\label{eq:propagation_jz}
  \frac{\partial}{\partial t} \j{z}(z,t) &= \Omega \j{y} +
  \frac{c\gamma}{8A\Delta}\frac{\lambda^2}{2\pi}a_2
       \left\{4\S{x}[\j{x}\j{y}+\j{y}\j{x}]-4\S{y}[\j{x}^2-\j{y}^2]\right\}.
\end{align}
The Larmor precession terms take all interesting dynamics to the
frequency component around $\Omega$. Let us see which terms above
couple $\Omega$-components of light into the spin state.
First, the $a_1\S{z}\j{x}$ term is our favorite interaction
term~(\ref{eq:jyout}) used in all quantum information protocols. It
consists of an atomic operator $\j{x}$ which is constant equal to $\pm
F$ for all practical purposes.  This is multiplied by $\S{z}$, whose
$\Omega$-components drive  $\partial
\hat{\jmath}_{y}/\partial t$.

In Eq.~(\ref{eq:propagation_jz}) the final term is proportional to
$\j{x}^2-\j{y}^2$. Practically, this is also constant equal to
$F^2-F/2$. It is then multiplied by $\S{y}$, whose
$\Omega$-components drive $\partial \hat{\jmath}_{z}/\partial t$. 
This is an unwanted
effect. The ratio of the unwanted to wanted noise is found by squaring
these contributions. We get
\begin{equation}
\label{eq:bad_noise_pileup}
  \frac{\text{Unwanted noise}}{\text{Wanted noise}}
    = 4(2F-1)^2\left(\frac{a_2}{a_1}\right)^2
    \frac{\text{Noise}(\S{y})}{\text{Noise}(\S{z})}. 
\end{equation}
For our typical values of detuning we have $a_2/a_1 \approx 10^{-2}$
and the above ratio becomes $\approx 0.02\cdot
\text{Noise}(\S{y})/\text{Noise}(\S{z})$ for $F=4$. If our laser beam
is polarized along the $x$- or $y$-axis with a clean linear
polarization, the noise of $\S{y}$ and $\S{z}$ at frequency $\Omega$
can be shot noise limited, i.e.~by quantum noise (amplitude
noise of the laser does drive into $\S{y}$ and $\S{z}$ in the
case of clean linear polarization). In this case the unwanted noise
only contributes $\approx2\%$ of the total noise pile up. But if we
choose arbitrary polarization directions in the $xy$-plane, the
$\S{y}$-component will have non-zero mean value, and the fluctuations
at $\Omega$ will essentially be the amplitude noise of the laser at
$\Omega$. In this case, to keep the last term of
Eq.~(\ref{eq:propagation_jz}) from accumulating extra noise,
one requires the laser \emph{intensity} to be shot noise limited at
$\Omega$ (which is a more difficult condition to meet than clean
linear polarization). We thus have one motivation for choosing the
laser to be polarized along the $x$- or $y$-direction and not in
between. We typically do this in our experiments and for this reason
we have to compensate the Stark splitting discussed in
Sec.~\ref{sec:polarization_and_Stark}. For a further discussion of the
different higher order terms in the interaction,
see~\citep{julsgaard_phd}. A more thorough discussion of quantum noise
with the higher order terms included is given
by~\citet{kupriyanov:05}.

\end{document}